# On The EU/Eurozone Stability

**Dimitris Sardelis**

Athens, Greece, February19, 2013

**Abstract**

The aim of the present article is to offer a strictly mathematical, statistical treatment of the current account balances in EU and in the Eurozone. Based on Eurostat data, an overview of the total and annual balances is first made for different collections among the EU countries. Then, using the *Mathematica* technical computing software, curve fitting is employed to determine the functions which best reflect how surpluses and deficits accumulate with time. It is shown that both EU and the Eurozone economies ultimately have to pass through a critical, turning point beyond which the accumulation of deficits exceeds the accumulation of surpluses thus marking a period of instability. An interval estimate at a ninety eight percent degree of confidence, yields that EU is found in a phase of instability since 2011 while the instability turning point for the Eurozone is bound to occur any year from 2015 to 2018.

# Introduction

The *current account balance of a country* summarizes within a specific time period, the final outcome of all economics transactions conducted by the country as a whole with the rest of the world. Dissociated from all the details and all the operations (social, political, economical, e.t.c) taken to produce it, the current account balance of a country stands as a most representative parameter for its economic status at any time. It is a well defined, clear trace along the country□s path throughout time.

In much the same way, *the current account balance of any collection of countries* (artificial or otherwise) within a specific time period, reflects at a snapshot the collection□s economic status and its relation with the rest of the world, dissociated both from all the proccesses taken to produce its balance and from all possible mutual relations among its constituent parts. The change of scale from a country to a collection of countries does not modify the balances definition but only the perspective to view it.

Subsequently, current account balances by their very same definition, can very well be and they should be considered as a legitimate terrain for strictly mathematical, statistical research where the requirement for self-consistent exposition and specific testable predictions leaves no room for any re-cycling of philosophical pre-conceptions and arbitrary political projections. The present article is intended to adopt this viewpoint for exploring the EU and the Eurozone current account balances.

After specifying the data base[**1**] and a few necessary (mainly notational) definitions to be used (**Sections 1 and 2**), my enquiry begins with an overview of total balances for all EU countries for the time period 1995 to 2011 (**Section 3**). The total balances of some essential country groupings are then highlighted (**Section 4**). As it is common practice to measure most economic quantities as percentages of GDP, the EU GDP distribution is next illustrated for the year 2011 (**Section 5**), supplemented with a section presenting how the EU GDP distribution and the total GDP of various country groupings, vary over the years (**Section 6**). The next section (**Section 7)** presents how the annual current acount balances of various country groupings vary with time.

Sections 3 to 7 are basically descriptive in character and they could be summarized as a *phenomenology* of balances and of the gross domestic products in EU. The next considerably longer section (**Section 8**) is extensively more *theoretical and inferential* in character as it deals with curve fitting the accumulated surpluses and deficits in the EU and the Eurozone ending up with best fit functions incorporated with interval single predictions. Section 8 also unveils a relative asymmetry between the accumulation of surpluses and the accumulation of deficits, bringing up an instability issue for both EU and the Eurozone. This asymmetry is explored in the final two sections (**Section 9 and 10**) yielding interval predictions for the times beyond which the total accumulated deficits in both EU and the Eurozone will start to exceed the total accumulated surpluses.

For the construction of all tables, graphs, curve fitting and the incorporated statistical model analysis involving predictions, the *Mathematica* computing software[**2**] has been used. *Mathematica* has also been used for the editing of this article. Regarding the style of exposition, I have tried to refrain from making comments of any kind. Whenever I felt that some figures look more significant than others, I merely listed short statements for them under the heading of observations. In other words, the information contained in the tables of this work (especially in the tables of Sections 3 to 7) speak for themselves for any one who cares to see them.

# 1. Data

The short descriptions under each data listed below are taken intact from the Eurostat tables.

- **Eurostat Data 1: GDP at current prices**

**Code:** nama_gdp_c

**Short Description**: GDP (gross domestic product) is an indicator for a nation's economic situation. It reflects the total value of all goods and services produced less the value of goods and services used for intermediate consumption in their production. Expressing GDP in PPS (purchasing power standards) eliminates differences in price levels between countries, and calculations on a per head basis allows for the comparison of economies significantly different in absolute size.

- **Eurostat Data 2: Current account balance in % of GDP-annual data**

  **Code:** tipsbp20

  **Short Description:** The balance of payments is the statistical statement that systematically summarizes, for a specific time period, the economic transactions of an economy with the rest of the world. The Balance of Payments is broken down into three broad sub-balances: the current account, the capital account, and the financial account. Current account is the major driver of net lending/net borrowing of an economy; it provides important information about the economic relations of a country with the rest of the world**.** It covers all transactions (other than those in financial items) that involve economic values and occur between resident and non-resident units. The EIP scoreboard indicator is expressed in percentage of GDP and calculated as: CAB%GDP=CAB*100/GDP. The indicator is based on the Balance of Payments data reported to Eurostat by the 27 EU Member States. Definitions are based on the IMF's Fifth Balance of Payments Manual (BPM5).

- **Eurostat Data 3: General government deficit/surplus**

  **Code:** tec00127

  **Short Description:** The general government deficit/surplus is defined in the Maastricht Treaty as general government net borrowing/lending according to the European System of Accounts (ESA95). It is the difference between the revenue and the expenditure of the general government sector. The government deficit data related to the EDP (EDP B.9) differs from the deficit according ESA95 (B.9) for the treatment of interest relating to swaps and forward rate agreements. The general government sector comprises the sub-sectors of central government, state government, local government and social security funds. The series are presented as a percentage of GDP and in millions of euro. GDP used as a denominator is the gross domestic product at current market prices.

# 2. Definitions

## EU Regions

- **Eurozone Countries**

  Austria, Belgium, Cyprus, Estonia, Finland, France, Germany, Greece, Ireland, Italy, Luxembourg, Malta, Netherlands, Portugal, Slovakia, Slovenia, Spain

- **Rest of EU (EU10)**

  **Countries with Currency Pegged to Euro**

  Bulgaria, Denmark, Latvia, Lithuania

  **Countries with Floating Currency**

  Czech Republic, Hungary, Poland, Romania, Sweden, UK

## Units

- **Time : t (**years after 1995), or **T (**the time period between 1995 to 2011)
- **Money :** billion euros or percentage of GDP

## Balances

- **All Balances :** [Surplus(+) / Deficit(−)]
- **Current Account Balance (CAB):** Annual ($\mathbf{[CAB]}_t$ ) or Total ( $\mathbf{[CAB]}_T$ ), Rank (**R**)
- **General Government Balance (GGB):** Annual ($\mathbf{[GGB]}_t$ ) or Total ( $\mathbf{[GGB]}_T$ ), Rank ( $\mathbf{R}_1$)
- **Private Sector Balance (PSB):** Annual ($\mathbf{[PSB]}_t$ ) or Total ( $\mathbf{[PSB]}_T$), Rank ($\mathbf{R}_2$)

★ **Note 1**: The ranks **R,** $\mathbf{R}_1$ and $\mathbf{R}_2$ correspond to the balances **CAB**, **GGB** and **PSB,** respectively, sorted according to magnitude.

★ **Note 2**: The notation for the annual and the total balances can be extended by specifying the country or region of reference written within a parenthesis, e.g., $\mathbf{[CAB(country)]}_t$.

★ **Note 3**: The notation introduced above can also be applied for the GDP of a country or a region at a particular year, e.g., $\mathbf{[GDP(region)]}_t$.

The annual or the total **CAB** of any EU country is found by combining **Eurostat Data 1** and **Eurostat Data 2.** The annual or the total **GGB** of any EU country is found directly from **Eurostat Data 3**. Finally, the annual or the total **PSB** is introduced here as the complement of **GGB,** i.e., the balance of all non-public economic transactions of a country with the rest of the world. Thus, for any time unit, the following basic account balance identity is taken to define **PSB**

$$\mathbf{(CAB) = (GGB) + (PSB)} \qquad \mathbf{(1)}$$

Consequently, **(1)** can be used to deduce **PSB** of any country from its **CAB** and **GGB** balances.

It must be remarked that all three balances are *algebraically additive* not only with respect to time (adding annual figures to produce balances for any specific time period ) but also with respect to countries: For any conceivable collection of countries, each one of the three balances, is the algebraic sum of the corresponding balances of its constituent countries. That this sum rule of balances holds true derives from the fact that the balance components of any two countries referring to their mutual/bilateral relation, cancel each other out.

# 3. Overview of Total Balances

The total balances of all EU countries for the time period 1995-2011 are diplayed in Table 1 below.

**Table 1. The Total Balances of the 27 EU Member States [Time Period:1995-2011]**

| COUNTRY | $[CAB]_T$ | R | $[GGB]_T$ | $R_1$ | $[PSB]_T$ | $R_2$ |
|---|---|---|---|---|---|---|
| Austria | 45.1827 | 7 | -94.2458 | 18 | 139.428 | 8 |
| Belgium | 119.102 | 4 | -83.0813 | 16 | 202.183 | 7 |
| Bulgaria⋆ | -31.7251 | 17 | -1.3856 | 6 | -30.3395 | 23 |
| Cyprus | -13.7309 | 13 | -6.5576 | 9 | -7.1733 | 19 |
| CzechRep | -56.6669 | 19 | -65.2647 | 15 | 8.59785 | 14 |
| Denmark | 91.0439 | 6 | 30.2217 | 2 | 60.8222 | 10 |
| Estonia | -11.191 | 12 | 0.5804 | 5 | -11.7714 | 22 |
| Finland | 104.578 | 5 | 42.486 | 1 | 62.0924 | 9 |
| France | 36.9069 | 9 | -1023.22 | 27 | 1060.13 | 2 |
| Germany | 1097.96 | 1 | -977.186 | 26 | 2075.15 | 1 |
| Greece⋆ | -228.18 | 25 | -181.853 | 21 | -46.3276 | 25 |
| Hungary | -61.3845 | 20 | -62.5755 | 14 | 1.19096 | 16 |
| Ireland | -30.4803 | 16 | -85.463 | 17 | 54.9827 | 11 |
| Italy | -143.072 | 22 | -798.345 | 24 | 655.273 | 3 |
| Latvia | -17.0055 | 14 | -6.3261 | 8 | -10.6794 | 21 |
| Lithuania | -21.3161 | 15 | -11.3564 | 10 | -9.95973 | 20 |
| Luxembourg | 43.1605 | 8 | 7.7302 | 4 | 35.4303 | 12 |
| Malta | -4.50554 | 10 | -3.9249 | 7 | -0.580644 | 17 |
| Netherlands | 468.602 | 2 | -151.113 | 20 | 619.715 | 4 |
| Poland | -171.858 | 23 | -185.981 | 22 | 14.1236 | 13 |
| Portugal | -206.848 | 24 | -118.337 | 19 | -88.511 | 26 |
| Romania | -89.0599 | 21 | -53.0349 | 13 | -36.025 | 24 |
| Slovakia | -32.8685 | 18 | -32.1526 | 12 | -0.71593 | 18 |
| Slovenia | -7.64566 | 11 | -14.6879 | 11 | 7.04224 | 15 |
| Spain | -673.525 | 27 | -430.834 | 23 | -242.691 | 27 |
| Sweden | 290.364 | 3 | 21.8175 | 3 | 268.546 | 6 |
| UK | -530.348 | 26 | -951.837 | 25 | 421.489 | 5 |
| EU27 | -34.5079 | | -5235.93 | | 5201.42 | |

⋆ **Note**: **Eurostat Data** (**2** and **3**) states the annual balances of all EU countries for the time period 1995-2011, except those of **Bulgaria** and **Greece** stated for the periods 1998-2011 and 2000-2011, respectively.

## Observations

- There are nine countries with a positive **CAB** sum total , i.e., a total current account *surplus*: **Austria, Belgium, Denmark, Finland, France, Germany, Luxembourg, Netherlands** and **Sweden**. The total surplus of **Germany** almost equals that of the other eight surplus EU countries combined.
  For the period 1995-2011, the total surplus distributed among the nine surplus countries is best illustrated by the following ***Surpluses PieChart*** (**Figure 1**) where the CAB totals have been ranked according to magnitude :

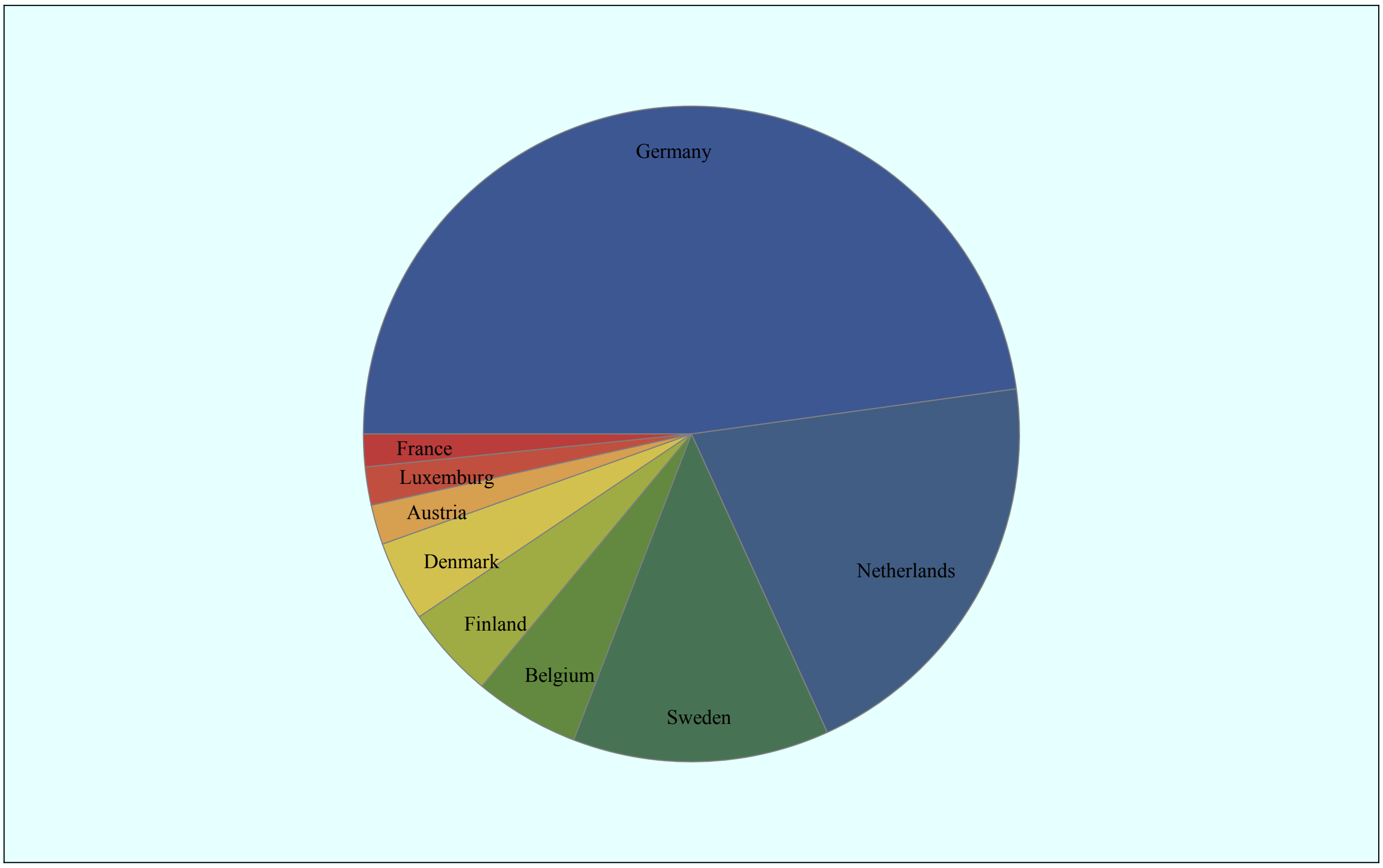

Figure 1: The EU Countries with a Total CAB Surplus [Time Period:1995-2011]

- On the opposite side, among the eighteen countries with a negative **CAB** sum total, i.e., a total current account *deficit*, the sum total deficit of **Spain** and **UK** is roughly fifty percent of the grand sum total deficit.
  For the period 1995-2011, the total deficit distributed among the eighteen deficit countries is best illustrated by the following ***Deficits PieChart*** (**Figure 2**) where the negative CAB totals have been ranked according to their *absolute* magnitude :

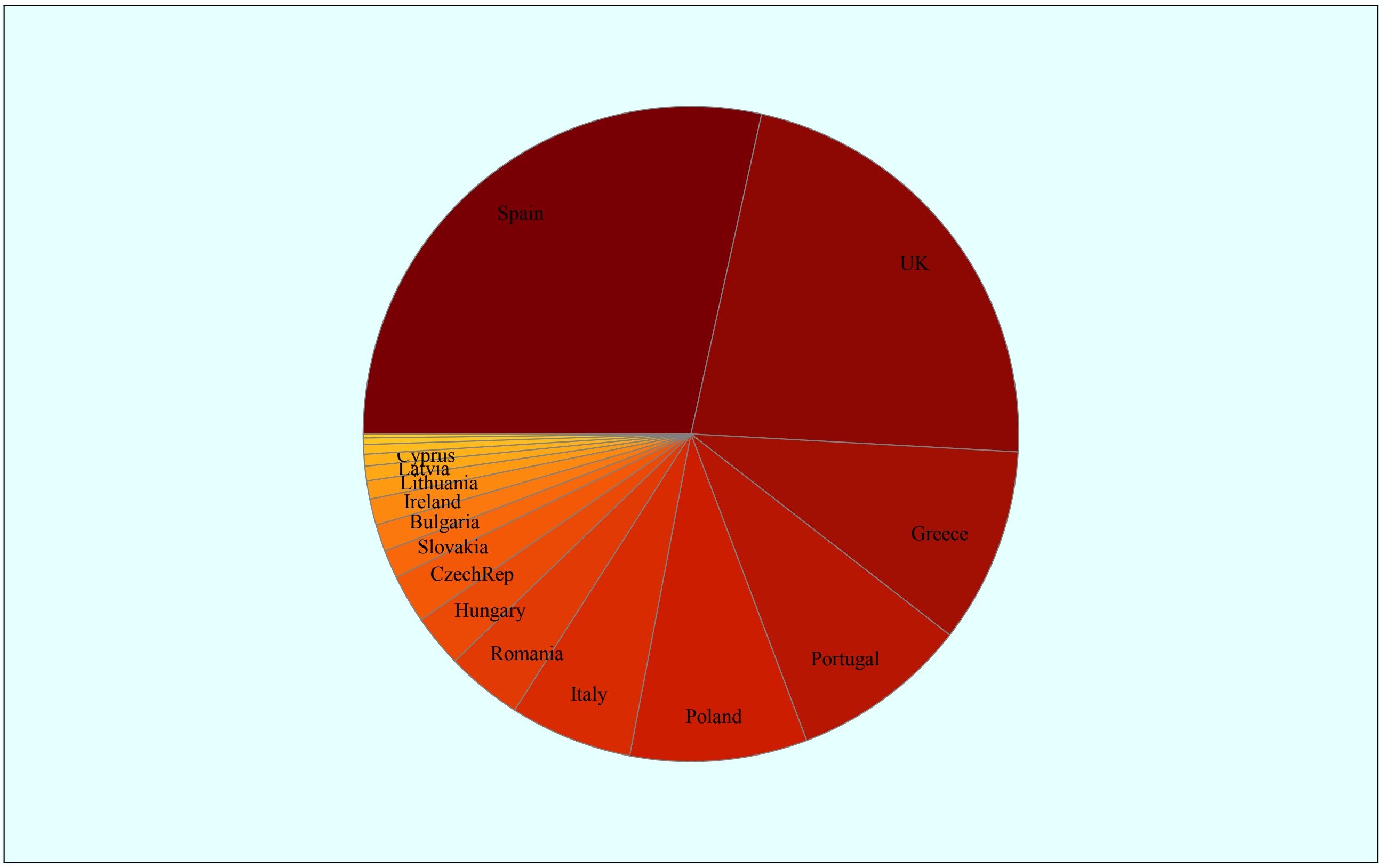

Figure 2: The EU Countries with a Total CAB Deficit [Time Period:1995-2011]

★ **Note**: The legends for **Estonia**, **Slovenia** and **Malta** are missed out from **Figure 2** as their shares in the deficit pie are hardly visible.

- **Germany** and **France** have the highest public debts, i.e., the most negative **GGB** sum totals, and also the highest **PSB** sum totals. The latter counterbalance the former yielding positive **CAB** sum totals, i.e., surpluses, for both countries.
- Four among the nine surplus countries, have kept up both the **GGB** and **PSB** sum totals positive: **Denmark, Finland, Luxembourg,** and **Sweden**.
- Ten countries have all three **CAB, GGB** and **PSB** sum totals negative: **Bulgaria, Cyprus, Greece, Latvia, Lithuania, Malta, Portugal, Romania, Slovakia** and **Spain.** The country with the highest deficit, i.e., with the most negative **CAB** sum total**,** is **Spain**.
- Next to **France** and **Germany, UK** has the third highest public debt, i.e., the third most negative **GGB** sum total and the fifth highest **PSB** sum total. These combined yield a negative **CAB** sum total, i.e., a total current account deficit, the second highest after **Spain**.

# 4. Aspects of Total Balances

The total balances may also be viewed by grouping the EU/Eurozone countries into different complementary regions and by adding algebraicaly the total balances of the constituent countries.

Thus, denoting as **EU9+** the nine surplus EU countries, **EU18-** the eighteen deficit EU countries, and **EU26** all EU countries except **Germany**, **Table 2** diplays the total balances of the following three paired/complementary regions: **Eurozone** versus **EU10**, **EU9+** versus **EU18-** and **Germany** versus **EU26**.

**Table 2. EU Total Balances in Complementary Pairs**

| REGIONS | $[CAB]_T$ | $[GGB]_T$ | $[PSB]_T$ |
|---|---|---|---|
| Eurozone | 563.448 | -3950.2 | 4513.65 |
| EU10 | -597.956 | -1285.72 | 687.767 |
| EU9+ | 2296.9 | -2226.59 | 4523.49 |
| EU18- | -2331.41 | -3009.34 | 677.926 |
| Germany | 1097.96 | -977.186 | 2075.15 |
| EU26 | -1132.47 | -4258.74 | 3126.27 |
| EU27 | -34.5079 | -5235.93 | 5201.42 |

Similarly, denoting as **EU7+** the seven Eurozone surplus countries, **EU10-** the ten deficit Eurozone countries, and **Eurozone16** all Eurozone countries except **Germany**, **Table 3** below diplays the total balances of two paired/complementary Eurozone regions: **Eurozone7+** versus **Eurozone10-** and **Germany** versus **Eurozone16**.

**Table 3. Eurozone Total Balances in Complementary Pairs**

| REGIONS | $[CAB]_T$ | $[GGB]_T$ | $[PSB]_T$ |
|---|---|---|---|
| Eurozone7+ | 1915.49 | -2278.63 | 4194.12 |
| Eurozone10- | -1352.05 | -1671.57 | 319.528 |
| Germany | 1097.96 | -977.186 | 2075.15 |
| Eurozone16 | -534.514 | -2973.02 | 2438.5 |
| Eurozone | 563.448 | -3950.2 | 4513.65 |

Finally, viewing the Eurozone in a more detailed surplus-deficit scale, we may also categorize the twenty seven EU Member States into the following four regions:

- **Germany**
- **Eurozone6 + :** Austria, Belgium, Finland, France, Luxembourg, Netherlands
- **Eurozone10 −:** Cyprus, Estonia, Greece, Ireland, Italy, Malta, Portugal, Slovakia, Slovenia, Spain
- **EU10:** Bulgaria, CzechRep, Denmark, Hungary, Latvia, Lithuania, Poland, Romania, Sweden, UK

The total balances of these four regions are displayed in **Table 4** below:

**Table 4. EU Total Balances in Four Regions**

| REGIONS | $[CAB]_T$ | $[GGB]_T$ | $[PSB]_T$ |
|---|---|---|---|
| Germany | 1097.96 | -977.186 | 2075.15 |
| Eurozone6+ | 817.532 | -1301.44 | 2118.98 |
| Eurozone10- | -1352.05 | -1671.57 | 319.528 |
| EU10 | -597.956 | -1285.72 | 687.767 |
| EU27 | -34.5079 | -5235.93 | 5201.42 |

## Observations

- The private sector total balance of the nine surplus EU countries outweighs the sum of their general government total balances in a ratio 2 to 1.
- On the contrary, the private sector total balance of the eighteen deficit EU countries is outweighed by the sum of their general government total balances in a ratio that is over 4 to 1.
- The total public deficit (**GGB**) of all EU countries is nearly equal to their total private sector net-saving (**PSB**) and the total current account balance (**CAB**) of all EU countries is close to zero.
- The total public deficit (**GGB**) of the Eurozone countries is less than their private sector net-saving (**PSB**) and their total current account balance (**CAB**) is positive (surplus).
- The private sector total balance of **Germany** outweighs its general government total balance in a ratio 2 to 1.
- The sum of the private sector total balances of the **Eurozone6+** countries outweighs that of their general government total balances in a ratio 11 to 7.
- On the contrary, the sum of the private sector total balances of the **Eurozone10−** countries is outweighed by the sum of their general government total balances in a ratio just over 5 to 1.

# 5. Overview of the EU Gross Domestic Product Distribution

Let us picture the GDP spectrum within EU at a particular year, e.g, 2011, the latest year with figures stated in **Eurostat Data 1**.The GDP distribution among the EU countries for 2011, is best illustrated by the following ***GDP PieChart*** (**Figure 3**) where the GDP of each EU country has been ranked according to magnitude :

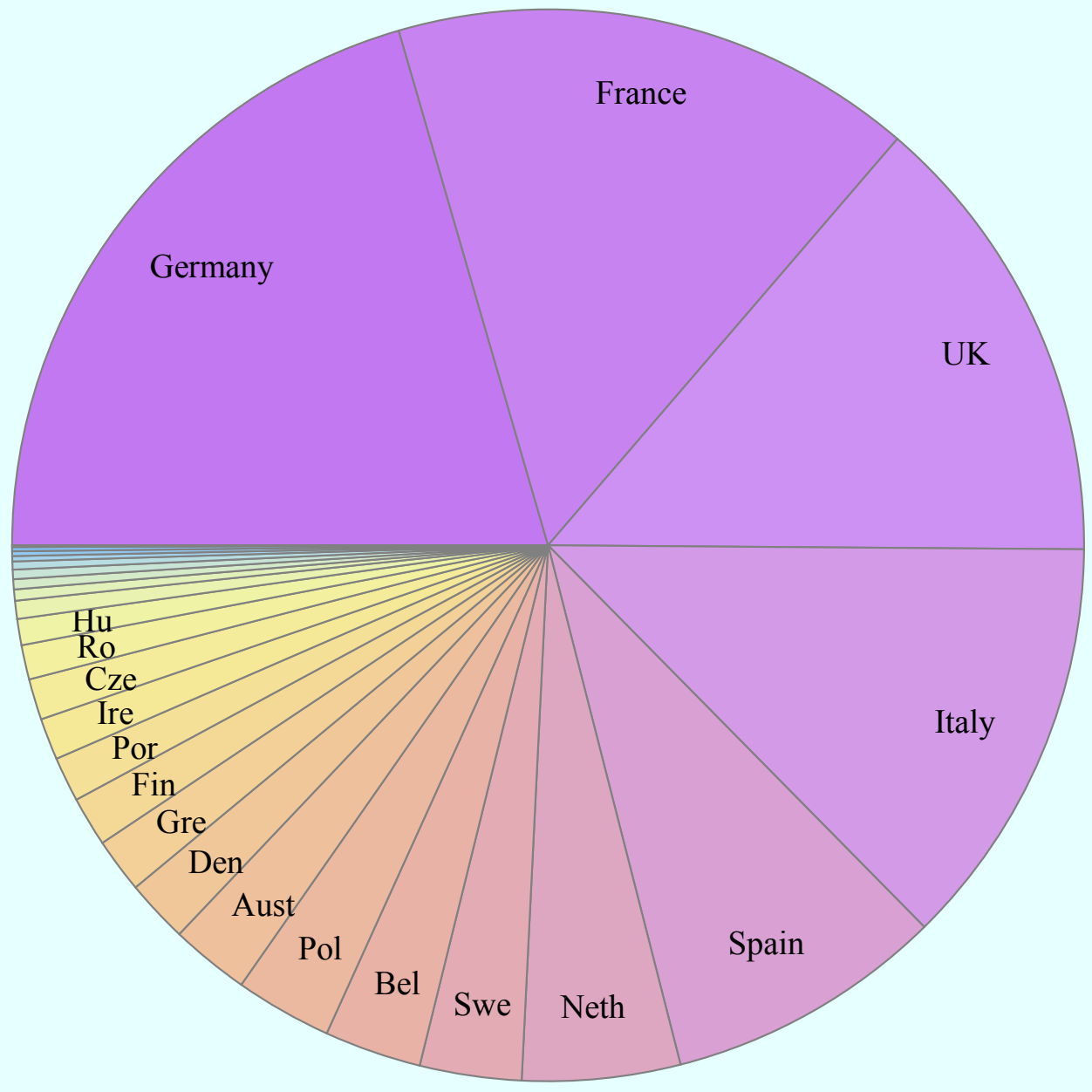

Figure 3: The GDP of the EU Countries [Year:2011]

★ **Note**: The legends for nine countries (**Slovakia, Luxembourg ,Bulgaria, Slovenia, Lithuania, Latvia, Cyprus, Estonia, Malta**) are missed out from **Figure 3** as their tiny shares in the wealth pie (total GDP of EU), are hardly visible.

## Observations

- Thus, for the year 2011, three countries (**Germany, France and UK**) own about fifty percent of the total EU wealth as reflected by the GDP distribution spectrum. Another twenty five percent belongs to three more countries (**Italy, Spain and Netherlands**) while the remaining twenty five percent is shared (also unevenly) by all the rest.

# 6. Aspects of the Time Changing GDP within EU

Neither the GDP pie size nor the GDP distribution among the EU countries stay fixed in time. A closer look at **Eurostat Data 1** unveils some noteworthy features shown in **Table 5** below:

**Table 5. $[GDP]_t/[GDP(E27)]_t$ for the Six EU Countries with Highest GDP**

| Year | t | Germany | France | UK | Italy | Spain | Netherlands | Total |
|---|---|---|---|---|---|---|---|---|
| 1995 | 0 | 0.27416 | 0.170865 | 0.127185 | 0.122979 | 0.0648636 | 0.0455403 | 0.805593 |
| 1996 | 1 | 0.259792 | 0.167632 | 0.131052 | 0.134934 | 0.0663498 | 0.0445486 | 0.804309 |
| 1997 | 2 | 0.243925 | 0.160718 | 0.154613 | 0.135541 | 0.0647431 | 0.0436975 | 0.803238 |
| 1998 | 3 | 0.238103 | 0.160581 | 0.15963 | 0.133814 | 0.065679 | 0.0440201 | 0.801828 |
| 1999 | 4 | 0.23289 | 0.159165 | 0.164285 | 0.132035 | 0.0675246 | 0.0449658 | 0.800866 |
| 2000 | 5 | 0.22253 | 0.156462 | 0.173917 | 0.130235 | 0.0684608 | 0.0454255 | 0.79703 |
| 2001 | 6 | 0.219313 | 0.156046 | 0.171113 | 0.131024 | 0.0709928 | 0.0467163 | 0.795205 |
| 2002 | 7 | 0.21461 | 0.155299 | 0.171042 | 0.131036 | 0.0734012 | 0.0468246 | 0.792212 |
| 2003 | 8 | 0.212536 | 0.157153 | 0.162557 | 0.132802 | 0.0775009 | 0.0472028 | 0.789752 |
| 2004 | 9 | 0.207026 | 0.156099 | 0.166701 | 0.131788 | 0.0793232 | 0.0463123 | 0.78725 |
| 2005 | 10 | 0.200898 | 0.155166 | 0.166777 | 0.129727 | 0.0821237 | 0.0463686 | 0.781061 |
| 2006 | 11 | 0.19775 | 0.15367 | 0.167125 | 0.127597 | 0.0842266 | 0.0461678 | 0.776537 |
| 2007 | 12 | 0.195747 | 0.152083 | 0.166325 | 0.125275 | 0.0848892 | 0.0460873 | 0.770407 |
| 2008 | 13 | 0.198331 | 0.154989 | 0.145079 | 0.126283 | 0.0872108 | 0.0476611 | 0.759554 |
| 2009 | 14 | 0.202009 | 0.16043 | 0.133861 | 0.129287 | 0.0891629 | 0.0487676 | 0.763518 |
| 2010 | 15 | 0.203275 | 0.157759 | 0.13922 | 0.126473 | 0.0854145 | 0.0479433 | 0.760085 |
| 2011 | 16 | 0.205066 | 0.157923 | 0.138192 | 0.124946 | 0.084108 | 0.0476142 | 0.75785 |

## Observations

- For the period 1995-2011, the portion of the EU GDP pie shared by the six countries with the highest GDP, has been reduced by almost five percent.
- For the same period, among these six countries, **Germany** has suffered a significant loss as its portion from the EU GDP pie decreased by almost seven percent. Next comes **France** with a reduction of its share by about one percent.
- For the same period, **UK** and **Spain** have increased their GDP pie share by almost one and two percent, respectively.
- Finally, **Italy** and **Netherlands** have retained their GDP pie share over the years, almost steadily.

**Table 6** below displays the GDP of some essential EU regions.

**Table 6. $[GDP]_t$ for the Main EU Regions**

| Year | t | EU9+ | EU18- | Germany | EU26 | Eurozone6+ | Eurozone10- | Eurozone | EU10 | EU27 |
|---|---|---|---|---|---|---|---|---|---|---|
| 1995 | 0 | 4301.42 | 2736.35 | 1929.47 | 5108.29 | 2038.9 | 1608.2 | 5576.57 | 1461.19 | 7037.76 |
| 1996 | 1 | 4366.57 | 3025.71 | 1920.45 | 5471.82 | 2083.06 | 1798.96 | 5802.48 | 1589.8 | 7392.28 |
| 1997 | 2 | 4402.78 | 3404.03 | 1904.28 | 5902.54 | 2124.65 | 1909.87 | 5938.81 | 1868.01 | 7806.82 |
| 1998 | 3 | 4553.55 | 3621.32 | 1946.47 | 6228.41 | 2224.8 | 1997.68 | 6168.94 | 2005.93 | 8174.87 |
| 1999 | 4 | 4739.74 | 3848.86 | 2000.2 | 6588.4 | 2333.55 | 2113.52 | 6447.27 | 2141.33 | 8588.6 |
| 2000 | 5 | 4962.12 | 4238.87 | 2047.5 | 7153.49 | 2472.77 | 2263.29 | 6783.56 | 2417.43 | 9200.99 |
| 2001 | 6 | 5114.02 | 4470.02 | 2101.9 | 7482.13 | 2579.15 | 2403.26 | 7084.31 | 2499.72 | 9584.03 |
| 2002 | 7 | 5248.6 | 4686.63 | 2132.2 | 7803.04 | 2664.92 | 2533.27 | 7330.39 | 2604.85 | 9935.24 |
| 2003 | 8 | 5352.27 | 4751.89 | 2147.5 | 7956.66 | 2737.35 | 2661.98 | 7546.84 | 2557.32 | 10104.2 |
| 2004 | 9 | 5536.86 | 5069.03 | 2195.7 | 8410.2 | 2852.46 | 2811.97 | 7860.13 | 2745.77 | 10605.9 |
| 2005 | 10 | 5697.95 | 5374.34 | 2224.4 | 8847.89 | 2967.83 | 2952.96 | 8145.19 | 2927.1 | 11072.3 |
| 2006 | 11 | 5966.69 | 5734.44 | 2313.9 | 9387.23 | 3115.87 | 3134.61 | 8564.38 | 3136.75 | 11701.1 |
| 2007 | 12 | 6279.7 | 6126.59 | 2428.5 | 9977.8 | 3285.73 | 3315.52 | 9029.75 | 3376.55 | 12406.3 |
| 2008 | 13 | 6422.03 | 6051.07 | 2473.8 | 9999.29 | 3379.84 | 3388.01 | 9241.65 | 3231.45 | 12473.1 |
| 2009 | 14 | 6174.82 | 5579.62 | 2374.5 | 9379.94 | 3284.27 | 3263.55 | 8922.32 | 2832.11 | 11754.4 |
| 2010 | 15 | 6469.85 | 5810.07 | 2496.2 | 9783.71 | 3387.22 | 3292.96 | 9176.38 | 3103.53 | 12279.9 |
| 2011 | 16 | 6721.75 | 5920.98 | 2592.6 | 10050.1 | 3501.1 | 3327.33 | 9421.03 | 3221.7 | 12642.7 |

The annual GDP for each region of **Table 6**, is expressed in **Table 7** as a fraction of the total GDP for any given year. The table manifests directly the GDP (re-)distribution within EU over time:

**Table 7. $[GDP]_t/[GDP(E27)]_t$ for the Main EU Regions**

| Year | t | EU9+ | EU18- | Germany | EU26 | Eurozone6+ | Eurozone10- | Eurozone | EU10 |
|---|---|---|---|---|---|---|---|---|---|
| 1995 | 0 | 0.611191 | 0.388809 | 0.27416 | 0.72584 | 0.289709 | 0.22851 | 0.792378 | 0.207622 |
| 1996 | 1 | 0.590694 | 0.409306 | 0.259792 | 0.740208 | 0.281789 | 0.243357 | 0.784938 | 0.215062 |
| 1997 | 2 | 0.563967 | 0.436033 | 0.243925 | 0.756075 | 0.272154 | 0.244642 | 0.760721 | 0.239279 |
| 1998 | 3 | 0.557018 | 0.442982 | 0.238103 | 0.761897 | 0.272151 | 0.244368 | 0.754622 | 0.245378 |
| 1999 | 4 | 0.551865 | 0.448135 | 0.23289 | 0.76711 | 0.271703 | 0.246085 | 0.750678 | 0.249322 |
| 2000 | 5 | 0.539303 | 0.460697 | 0.22253 | 0.77747 | 0.268751 | 0.245983 | 0.737264 | 0.262736 |
| 2001 | 6 | 0.533598 | 0.466402 | 0.219313 | 0.780687 | 0.269109 | 0.250757 | 0.739178 | 0.260822 |
| 2002 | 7 | 0.528282 | 0.471718 | 0.21461 | 0.78539 | 0.268229 | 0.254978 | 0.737817 | 0.262183 |
| 2003 | 8 | 0.529709 | 0.470291 | 0.212536 | 0.787464 | 0.270914 | 0.263454 | 0.746904 | 0.253096 |
| 2004 | 9 | 0.522055 | 0.477945 | 0.207026 | 0.792974 | 0.26895 | 0.265132 | 0.741109 | 0.258891 |
| 2005 | 10 | 0.514614 | 0.485386 | 0.200898 | 0.799102 | 0.268041 | 0.266699 | 0.735638 | 0.264362 |
| 2006 | 11 | 0.509924 | 0.490076 | 0.19775 | 0.80225 | 0.266288 | 0.267889 | 0.731927 | 0.268073 |
| 2007 | 12 | 0.506171 | 0.493829 | 0.195747 | 0.804253 | 0.264843 | 0.267245 | 0.727836 | 0.272164 |
| 2008 | 13 | 0.51487 | 0.48513 | 0.198331 | 0.801669 | 0.27097 | 0.271625 | 0.740927 | 0.259073 |
| 2009 | 14 | 0.525318 | 0.474682 | 0.202009 | 0.797991 | 0.279407 | 0.277644 | 0.75906 | 0.24094 |
| 2010 | 15 | 0.526864 | 0.473136 | 0.203275 | 0.796725 | 0.275835 | 0.268158 | 0.747268 | 0.252732 |
| 2011 | 16 | 0.531669 | 0.468331 | 0.205066 | 0.794934 | 0.276926 | 0.263182 | 0.745174 | 0.254826 |

## Observations

- For the period 1995-2011, the portion of the EU GDP pie shared by the nine surplus countries, has been reduced by eight percent.
- Almost five percent of the EU GDP pie passed over to the Europe10 group (excepting Denmark and Sweden) and three percent to the Eurozone10− group.

# 7. Aspects of the Time Changing Current Account Balances

This section presents the annual current account balances for the country groupings introduced in Section 4: the four EU regions (**Tables 8 and 9**), the nine surplus versus the eighteen deficit EU countries (**Table 10**), Germany versus all other EU countries (**Table 11**) and, finally, the seven surplus versus the ten deficit Eurozone countries (**Table 12**).

## $[CAB]_t$ for the Four EU Regions

**Table 8. $[CAB]_t$ in billion euros of the Four EU Regions**

| Year | t | Germany | Eurozone6+ | Eurozone10- | EU10 | EU27 |
|---|---|---|---|---|---|---|
| 1995 | 0 | -23.1537 | 41.1819 | 18.5691 | -10.1808 | 26.4165 |
| 1996 | 1 | -11.5227 | 44.2447 | 24.9731 | -8.08727 | 49.6077 |
| 1997 | 2 | -9.5214 | 71.2742 | 22.1109 | -3.27294 | 80.5908 |
| 1998 | 3 | -13.6253 | 65.0771 | 3.70961 | -11.6536 | 43.5078 |
| 1999 | 4 | -26.0026 | 67.5522 | -17.8883 | -39.6397 | -15.9785 |
| 2000 | 5 | -34.8075 | 47.4918 | -54.6495 | -49.7098 | -91.675 |
| 2001 | 6 | 0. | 57.8715 | -50.6654 | -34.7341 | -27.528 |
| 2002 | 7 | 42.644 | 60.295 | -55.2277 | -30.3951 | 17.3163 |
| 2003 | 8 | 40.8025 | 54.8749 | -61.8718 | -22.394 | 11.4116 |
| 2004 | 9 | 103.198 | 72.799 | -77.3392 | -43.3996 | 55.2581 |
| 2005 | 10 | 113.444 | 49.6995 | -122.593 | -44.7888 | -4.23757 |
| 2006 | 11 | 145.776 | 63.7915 | -166.143 | -73.5118 | -30.0875 |
| 2007 | 12 | 179.709 | 46.9319 | -194.743 | -83.3019 | -51.4037 |
| 2008 | 13 | 153.376 | 8.89544 | -227.491 | -54.5398 | -119.759 |
| 2009 | 14 | 140.096 | 13.6741 | -132.323 | -17.6111 | 3.83552 |
| 2010 | 15 | 149.772 | 36.7949 | -143.814 | -47.3627 | -4.60989 |
| 2011 | 16 | 147.778 | 15.0827 | -116.661 | -23.3729 | 22.8273 |

### Observations

- Since 1999 (marking the beginning of EMU), the **Eurozone10 –** countries have been trapped in a phase of persistently increasing current account deficits which systematically exceed the deficits of the **Europe10** countries.
- **Germany** and, to a much lesser degree, the **Eurozone6+** countries, are the exclusive beneficiaries in the European economy: Their current account surpluses match, almost euro-for-euro, the current account deficits of the **Eurozone10 –** and the **Europe10** countries.

**Table 9** below complements **Table 8** displaying the annual current accounts of the four EU regions as fractions of their corresponding **GDP**.

**Table 9. $[CAB]_t / [GDP]_t$ for the Four EU Regions**

| Year | t | Germany | Eurozone6+ | Eurozone10- | EU10 | EU27 |
|---|---|---|---|---|---|---|
| 1995 | 0 | -0.012 | 0.0201981 | 0.0115465 | -0.0069675 | 0.00375353 |
| 1996 | 1 | -0.006 | 0.0212402 | 0.0138819 | -0.00508697 | 0.00671075 |
| 1997 | 2 | -0.005 | 0.0335463 | 0.0115771 | -0.0017521 | 0.0103231 |
| 1998 | 3 | -0.007 | 0.0292508 | 0.00185696 | -0.00580956 | 0.00532214 |
| 1999 | 4 | -0.013 | 0.0289483 | -0.00846374 | -0.0185117 | -0.00186043 |
| 2000 | 5 | -0.017 | 0.0192059 | -0.0241461 | -0.0205631 | -0.00996359 |
| 2001 | 6 | 0. | 0.0224382 | -0.0210819 | -0.0138952 | -0.00287228 |
| 2002 | 7 | 0.02 | 0.0226255 | -0.0218009 | -0.0116686 | 0.00174291 |
| 2003 | 8 | 0.019 | 0.0200467 | -0.0232427 | -0.00875682 | 0.0011294 |
| 2004 | 9 | 0.047 | 0.0255215 | -0.0275036 | -0.015806 | 0.00521013 |
| 2005 | 10 | 0.051 | 0.0167461 | -0.0415151 | -0.0153014 | -0.000382719 |
| 2006 | 11 | 0.063 | 0.0204731 | -0.0530029 | -0.0234356 | -0.00257134 |
| 2007 | 12 | 0.074 | 0.0142836 | -0.0587367 | -0.0246707 | -0.00414336 |
| 2008 | 13 | 0.062 | 0.00263191 | -0.0671458 | -0.0168778 | -0.00960141 |
| 2009 | 14 | 0.059 | 0.00416352 | -0.0405457 | -0.00621837 | 0.000326304 |
| 2010 | 15 | 0.06 | 0.0108628 | -0.0436732 | -0.0152609 | -0.000375401 |
| 2011 | 16 | 0.057 | 0.00430801 | -0.0350613 | -0.00725485 | 0.00180556 |

### Observations

- The current account deficit of the **Eurozone10 –** countries as a fraction of their total **GDP** increases with time.
- The current account deficit of the **Europe10** countries as a fraction of their total **GDP** is roughly constant.
- Although the current account surplus in billion euros of the **Eurozone6+** countries increases with time, expressed as a fraction of the **Eurozone6+** total **GDP**, it decreases with time almost steadily.

Thus, overall, it looks as if EMU was chiefly designed so that **Germany** drains the economic life out of the Eurozone and ultimately of the entire EU.

## $[\text{CAB}]_t$ for the Surplus versus the Deficit EU Countries

**Table 10. $[\text{CAB}]_t$ and $[\text{CAB}]_t/[\text{GDP}]_t$ for the Surplus and the Deficit EU Countries**

| Year | t | $[\text{CAB}(9+)]_t$ | $\frac{[\text{CAB}(9+)]_t}{[\text{GDP}(9+)]_t}$ | $[\text{CAB}(18-)]_t$ | $\frac{[\text{CAB}(18-)]_t}{[\text{GDP}(18-)]_t}$ | $[\text{CAB}(\text{EU27})]_t$ | $\frac{[\text{CAB}(\text{EU27})]_t}{[\text{GDP}(\text{EU27})]_t}$ |
|---|---|---|---|---|---|---|---|
| 1995 | 0 | 22.6865 | 0.00527419 | 3.73 | 0.00136313 | 26.4165 | 0.00375353 |
| 1996 | 1 | 42.377 | 0.00970488 | 7.23069 | 0.00238975 | 49.6077 | 0.00671075 |
| 1997 | 2 | 71.8162 | 0.0163115 | 8.77456 | 0.0025777 | 80.5908 | 0.0103231 |
| 1998 | 3 | 58.6861 | 0.012888 | -15.1782 | -0.00419135 | 43.5078 | 0.00532214 |
| 1999 | 4 | 54.605 | 0.0115207 | -70.5834 | -0.0183388 | -15.9785 | -0.00186043 |
| 2000 | 5 | 26.3813 | 0.00531654 | -118.056 | -0.0278509 | -91.675 | -0.00996359 |
| 2001 | 6 | 76.1146 | 0.0148835 | -103.643 | -0.0231862 | -27.528 | -0.00287228 |
| 2002 | 7 | 120.094 | 0.0228812 | -102.778 | -0.02193 | 17.3163 | 0.00174291 |
| 2003 | 8 | 121.332 | 0.0226692 | -109.92 | -0.0231318 | 11.4116 | 0.0011294 |
| 2004 | 9 | 201.157 | 0.0363305 | -145.899 | -0.0287824 | 55.2581 | 0.00521013 |
| 2005 | 10 | 192.349 | 0.0337575 | -196.586 | -0.0365787 | -4.23757 | -0.000382719 |
| 2006 | 11 | 242.856 | 0.0407019 | -272.944 | -0.0475973 | -30.0875 | -0.00257134 |
| 2007 | 12 | 260.579 | 0.0414955 | -311.983 | -0.0509228 | -51.4037 | -0.00414336 |
| 2008 | 13 | 199.416 | 0.0310519 | -319.175 | -0.052747 | -119.759 | -0.00960141 |
| 2009 | 14 | 180.967 | 0.0293072 | -177.131 | -0.0317461 | 3.83552 | 0.000326304 |
| 2010 | 15 | 223.965 | 0.0346168 | -228.575 | -0.0393412 | -4.60989 | -0.000375401 |
| 2011 | 16 | 201.52 | 0.0299803 | -178.693 | -0.0301796 | 22.8273 | 0.00180556 |

### Observations

- The current account balance of the nine surplus countries, expressed as a fraction of their corresponding total GDP year by year, increases with time at an average rate of 0.1544 percent annually.
- The annual total surpluses of the nine (surplus) countries almost counterbalance the annual total deficits of the eighteen (deficit) countries. This is graphically displayed in **Figure 4** below:

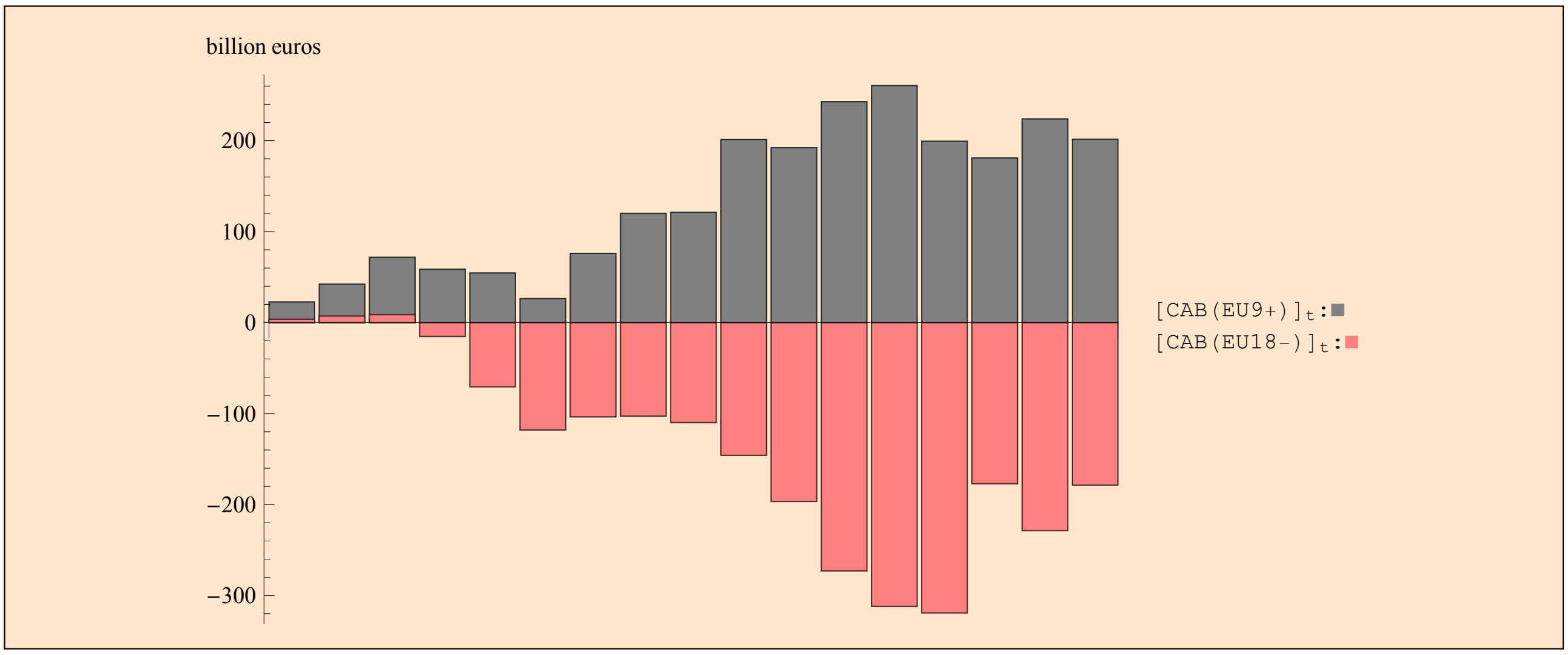

Figure 4: $[CAB]_t$ of the Surplus and of the Deficit Countries

# $[CAB]_t$ for Germany versus all other EU Countries

**Table 11. $[CAB]_t$ and $[CAB]_t/[GDP]_t$ for Germany and the rest of EU Countries**

| Year | t | $[CAB(Ger)]_t$ | $\frac{[CAB(Ger)]_t}{[GDP(Ger)]_t}$ | $[CAB(EU26)]_t$ | $\frac{[CAB(EU26)]_t}{[GDP(EU26)]_t}$ | $[CAB(EU27)]_t$ | $\frac{[CAB(EU27)]_t}{[GDP(EU27)]_t}$ |
|---|---|---|---|---|---|---|---|
| 1995 | 0 | -23.1537 | -0.012 | 49.5702 | 0.00970387 | 26.4165 | 0.00375353 |
| 1996 | 1 | -11.5227 | -0.006 | 61.1305 | 0.0111719 | 49.6077 | 0.00671075 |
| 1997 | 2 | -9.5214 | -0.005 | 90.1122 | 0.0152667 | 80.5908 | 0.0103231 |
| 1998 | 3 | -13.6253 | -0.007 | 57.1331 | 0.00917299 | 43.5078 | 0.00532214 |
| 1999 | 4 | -26.0026 | -0.013 | 10.0241 | 0.00152148 | -15.9785 | -0.00186043 |
| 2000 | 5 | -34.8075 | -0.017 | -56.8675 | -0.00794961 | -91.675 | -0.00996359 |
| 2001 | 6 | 0. | 0. | -27.528 | -0.00367917 | -27.528 | -0.00287228 |
| 2002 | 7 | 42.644 | 0.02 | -25.3277 | -0.00324588 | 17.3163 | 0.00174291 |
| 2003 | 8 | 40.8025 | 0.019 | -29.3909 | -0.00369387 | 11.4116 | 0.0011294 |
| 2004 | 9 | 103.198 | 0.047 | -47.9398 | -0.0057002 | 55.2581 | 0.00521013 |
| 2005 | 10 | 113.444 | 0.051 | -117.682 | -0.0133006 | -4.23757 | -0.000382719 |
| 2006 | 11 | 145.776 | 0.063 | -175.863 | -0.0187343 | -30.0875 | -0.00257134 |
| 2007 | 12 | 179.709 | 0.074 | -231.113 | -0.0231627 | -51.4037 | -0.00414336 |
| 2008 | 13 | 153.376 | 0.062 | -273.135 | -0.0273154 | -119.759 | -0.00960141 |
| 2009 | 14 | 140.096 | 0.059 | -136.26 | -0.0145268 | 3.83552 | 0.000326304 |
| 2010 | 15 | 149.772 | 0.06 | -154.382 | -0.0157795 | -4.60989 | -0.000375401 |
| 2011 | 16 | 147.778 | 0.057 | -124.951 | -0.0124328 | 22.8273 | 0.00180556 |

## Observations

- The current account balance of **Germany**, expressed as a fraction of its GDP year by year,  increases with time at an average rate of 0.4313 percent annually, i.e., almost three times as fast as that of the nine surplus countries
- The annual current account balances of **Germany** almost counterbalance the annual deficits of the rest of Europe. This is graphically displayed in **Figure 5** below:

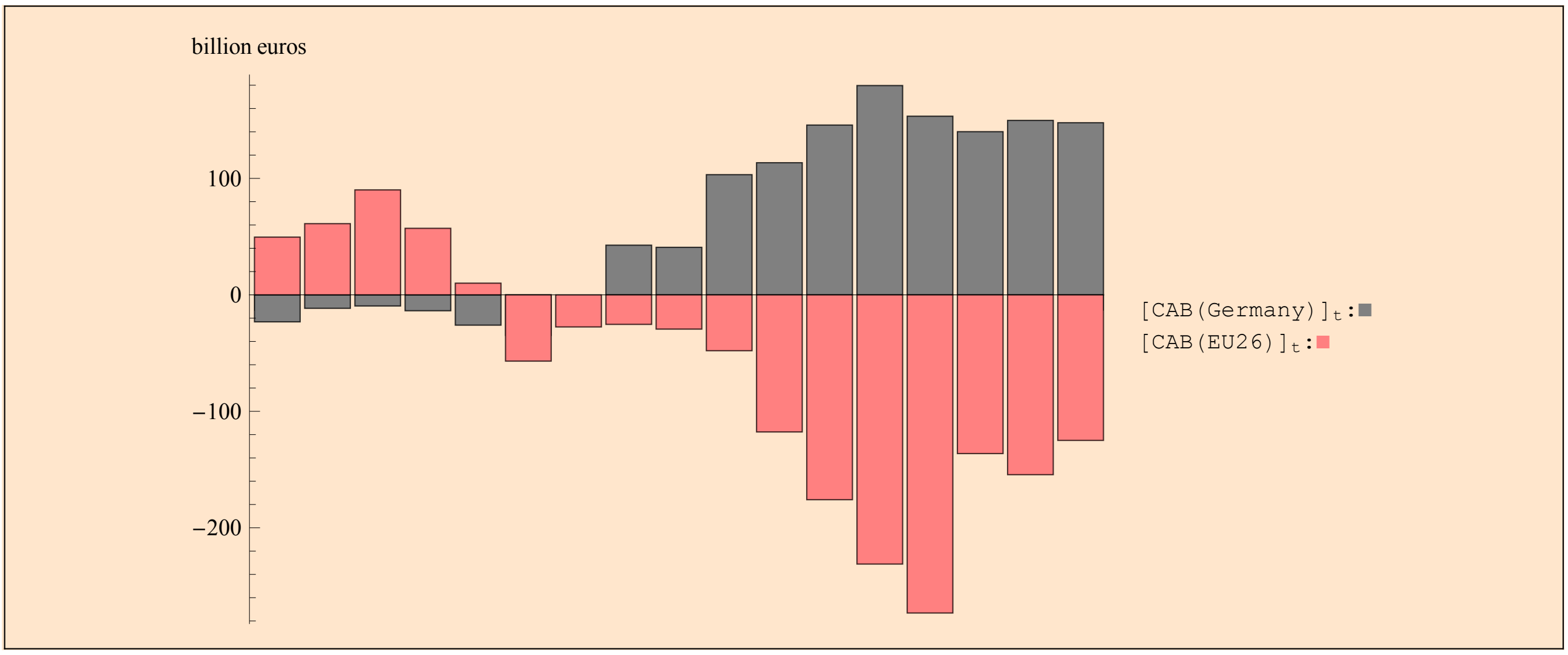

Figure 5: $[CAB]_t$ of Germany and of EU26

## $[CAB]_t$ for the Surplus versus the Deficit Eurozone Countries

**Table 12. $[CAB]_t$ and $[CAB]_t/[GDP]_t$ for the Surplus and the Deficit Eurozone Countries**

| Year | t | $[CAB(7+)]_t$ | $\frac{[CAB\ (Euro7+)]_t}{[GDP\ (Euro7+)]_t}$ | $[CAB(Euro10-)]_t$ | $\frac{[CAB\ (Euro10-)]_t}{[GDP\ (Euro10-)]_t}$ | $[CAB(Euro17)]_t$ | $\frac{[CAB\ (Euro17)]_t}{[GDP\ (Euro17)]_t}$ |
|---|---|---|---|---|---|---|---|
| 1995 | 0 | 18.0282 | 0.00454298 | 18.5691 | 0.0115465 | 36.5973 | 0.00656269 |
| 1996 | 1 | 32.7219 | 0.0081733 | 24.9731 | 0.0138819 | 57.695 | 0.00994317 |
| 1997 | 2 | 61.7528 | 0.0153273 | 22.1109 | 0.0115771 | 83.8637 | 0.0141213 |
| 1998 | 3 | 51.4518 | 0.0123348 | 3.70961 | 0.00185696 | 55.1614 | 0.0089418 |
| 1999 | 4 | 41.5496 | 0.00958745 | -17.8883 | -0.00846374 | 23.6613 | 0.00366997 |
| 2000 | 5 | 12.6843 | 0.0028061 | -54.6495 | -0.0241461 | -41.9651 | -0.0061863 |
| 2001 | 6 | 57.8715 | 0.0123629 | -50.6654 | -0.0210819 | 7.20609 | 0.00101719 |
| 2002 | 7 | 102.939 | 0.0214585 | -55.2277 | -0.0218009 | 47.7113 | 0.0065087 |
| 2003 | 8 | 95.6774 | 0.0195866 | -61.8718 | -0.0232427 | 33.8057 | 0.00447945 |
| 2004 | 9 | 175.997 | 0.0348636 | -77.3392 | -0.0275036 | 98.6576 | 0.0125517 |
| 2005 | 10 | 163.144 | 0.0314208 | -122.593 | -0.0415151 | 40.5512 | 0.00497855 |
| 2006 | 11 | 209.567 | 0.0385959 | -166.143 | -0.0530029 | 43.4242 | 0.00507033 |
| 2007 | 12 | 226.641 | 0.0396626 | -194.743 | -0.0587367 | 31.8981 | 0.00353256 |
| 2008 | 13 | 162.271 | 0.0277214 | -227.491 | -0.0671458 | -65.2195 | -0.00705713 |
| 2009 | 14 | 153.77 | 0.0271737 | -132.323 | -0.0405457 | 21.4466 | 0.0024037 |
| 2010 | 15 | 186.567 | 0.0317106 | -143.814 | -0.0436732 | 42.7528 | 0.00465901 |
| 2011 | 16 | 162.861 | 0.0267261 | -116.661 | -0.0350613 | 46.2002 | 0.00490394 |

### Observations

- The current account balance of the seven surplus Eurozone countries, expressed as a fraction of their corresponding total GDP year by year, increases with time at an average rate of 0.1304 percent annually.
- The annual total surpluses of the seven (surplus) countries exceed the annual total deficits of the ten (deficit) Eurozone countries. This is graphically displayed in **Figure 6** below:

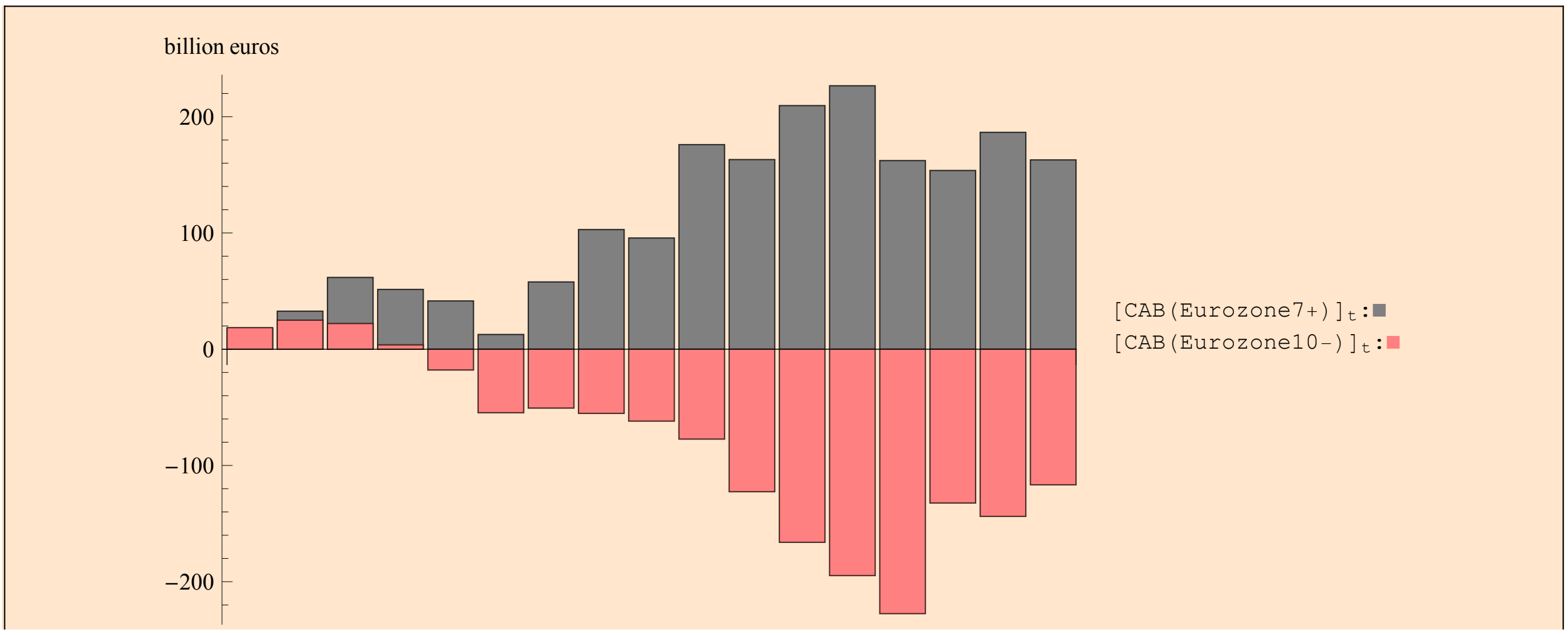

Figure 6: $[CAB]_t$ of the Surplus and of the Deficit EurozoneCountries

- The net annual current account balance of the entire Eurozone expressed as a fraction of the Eurozone total GDP year by year, is almost constant.
- Nevertheless, expressed in billion euros, it is persistently positive (the only two exceptions being the years 2000 and 2008). This is graphically displayed in **Figure 7** below:

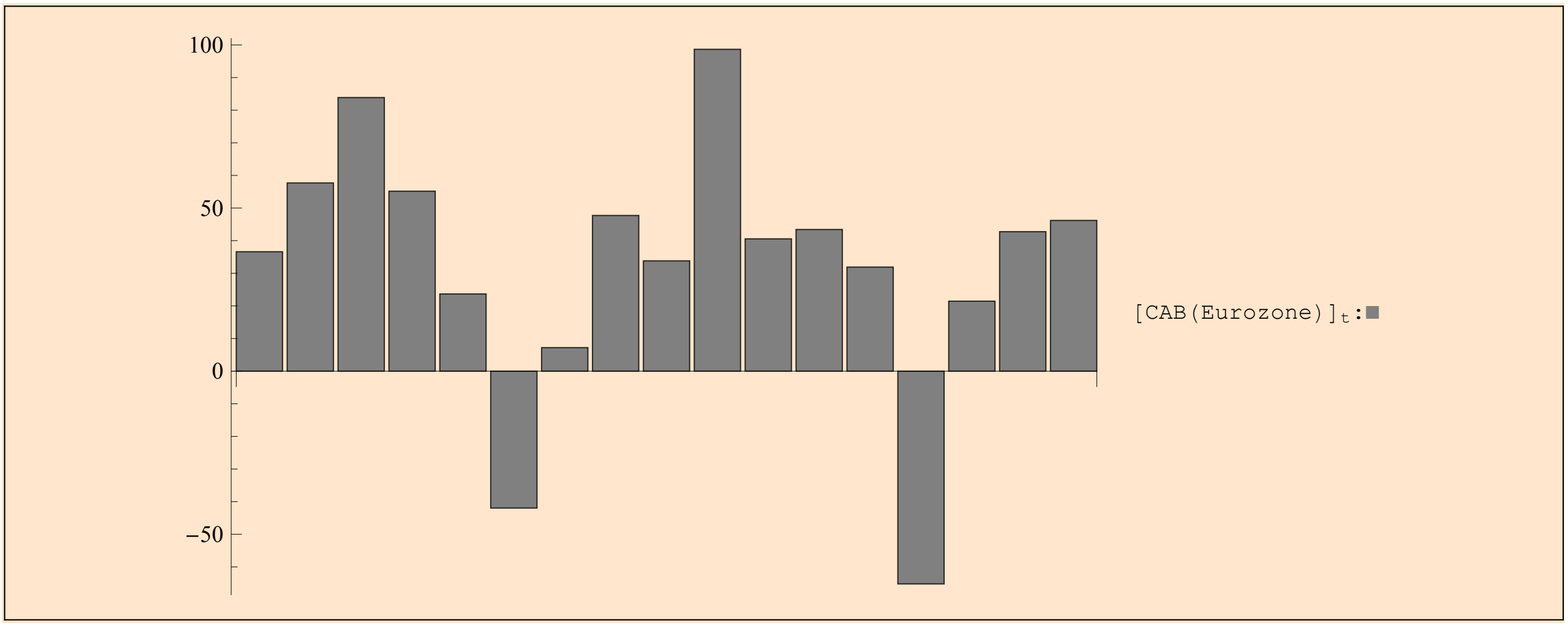

Figure 7: The Eurozone $[CAB]_t$

# 8. Curve Fitting the Accumulated Surpluses and Deficits

In this section we shall first determine the functional models which best describe (i) how the current account surplus of the nine surplus EU countries accumulates with time and (ii) how the current account deficit of the eighteen deficit EU countries accumulates with time. This will be achieved by means of exponential curve fitting the accumulated data of surpluses and of deficits of the two complementary EU regions. Following the same methodology, we shall next determine the best fit functional models for the accumulation of surpluses and for the accumulation of deficits of the seven Eurozone surplus countries and the ten Eurozone deficit countries, respectively. The comparison of the derived best fit functions will then be used to explore the EU/Eurozone stability issue.

## Curve Fitting Essentials using *Mathematica*

In order to determine the function that best fits any data to be considered below, we shall be applying the non-linear fit option for the case of exponential functions of the form $\alpha e^{\beta t}$ where $\alpha$ and $\beta$ are data dependent parameters. For every case considered, our exposition will consist of the following five parts: (I) **A statistical model analysis summary** reporting the best fit $\alpha$ and $\beta$ parameters with their corresponding ninety five per cent confidence intervals, supplemented by an ANOVA table which gives a formatted analysis of variance table for the model. (II) The **best fit function form** taken from the statistical model analysis summary. (III) **The determination coefficient($R^2$)** that expresses the model accuracy on a 0 to 1 scale. $R^2$ is defined from the ANOVA table as the ratio of the difference between the uncorrected total sum of squares and the residual sum of squares to the uncorrected total sum of squares. (IV) A single prediction confidence interval table, entitled as **Prediction** in short, that includes the observed and predicted responses, standard error estimates, and ninety five per cent confidence intervals for each given value of the independent time variable. In all cases to be considered, the uncertainty of the predictions increases with time as reflected in the time expanding confidence interval bands which incorporate both the variation in parameter estimates and the overall variation in response values. (V) **A graph** where the best fit function and its ninety five per cent confidence bands are superimposed on the original data which they were designed to reflect.

## Curve Fitting the Surplus Accumulation of the EU9+ Countries

### Statistical Model Analysis Summary

| | Estimate | Standard Error | Confidence Interval |
|---|---|---|---|
| $\alpha$ | 168.249 | 23.3499 | {118.48, 218.018} |
| $\beta$ | 0.169098 | 0.0100194 | {0.147742, 0.190454} |

| | DF | SS | MS |
|---|---|---|---|
| Model | 2 | $2.20164\times10^7$ | $1.10082\times10^7$ |
| Error | 15 | 256399. | 17093.2 |
| Uncorrected Total | 17 | $2.22728\times10^7$ | |
| Corrected Total | 16 | $9.34465\times10^6$ | |

### Best Fit Function Form

$$S_E(t) = 168.249\, e^{0.169098\, t} \qquad (2)$$

### Model Accuracy

**Determination Coefficient : 0.988488**

### Prediction

**Table 13 . [CAB(EU9 +)]$_T$ versus $S_E$(t) for $T$ = {0, 1, 2, ..., $t$}**

| Year | t | Observed | Predicted | SE | CILow | CIHigh |
|---|---|---|---|---|---|---|
| 1995 | 0 | 22.6865 | 168.249 | 132.81 | -114.828 | 451.327 |
| 1996 | 1 | 65.0635 | 199.247 | 133.243 | -84.7541 | 483.247 |
| 1997 | 2 | 136.88 | 235.955 | 133.733 | -49.0905 | 521. |
| 1998 | 3 | 195.566 | 279.426 | 134.274 | -6.77159 | 565.624 |
| 1999 | 4 | 250.171 | 330.907 | 134.851 | 43.4782 | 618.335 |
| 2000 | 5 | 276.552 | 391.872 | 135.442 | 103.184 | 680.559 |
| 2001 | 6 | 352.667 | 464.069 | 136.011 | 174.167 | 753.97 |
| 2002 | 7 | 472.761 | 549.567 | 136.514 | 258.593 | 840.54 |
| 2003 | 8 | 594.093 | 650.817 | 136.898 | 359.026 | 942.607 |
| 2004 | 9 | 795.249 | 770.72 | 137.111 | 478.475 | 1062.97 |
| 2005 | 10 | 987.598 | 912.715 | 137.133 | 620.423 | 1205.01 |
| 2006 | 11 | 1230.45 | 1080.87 | 137.02 | 788.818 | 1372.92 |
| 2007 | 12 | 1491.03 | 1280. | 137.005 | 987.986 | 1572.02 |
| 2008 | 13 | 1690.45 | 1515.83 | 137.666 | 1222.4 | 1809.26 |
| 2009 | 14 | 1871.42 | 1795.1 | 140.196 | 1496.28 | 2093.92 |
| 2010 | 15 | 2095.38 | 2125.82 | 146.732 | 1813.07 | 2438.57 |
| 2011 | 16 | 2296.9 | 2517.47 | 160.539 | 2175.29 | 2859.65 |
| 2012 | 17 | - | 2981.28 | 185.729 | 2585.41 | 3377.15 |
| 2013 | 18 | - | 3530.54 | 226.572 | 3047.61 | 4013.47 |
| 2014 | 19 | - | 4180.99 | 287.088 | 3569.08 | 4792.9 |
| 2015 | 20 | - | 4951.28 | 371.347 | 4159.77 | 5742.79 |

### Graph

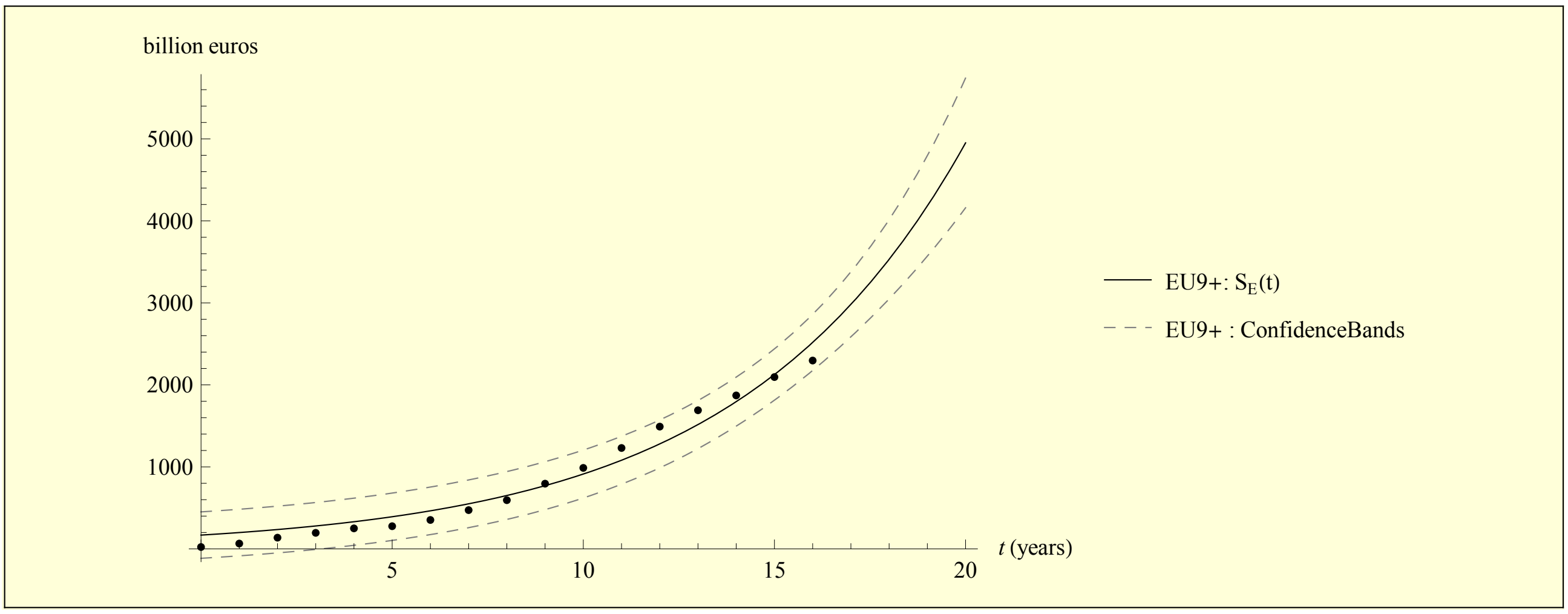

Figure 8: The **[CAB(EU9 +)]$_T$** DataPlot and the ***$S_E(t)$*** Graph with 95 per cent Confidence Bands

## Curve Fitting the Deficit Accumulation of the EU18- Countries

### Statistical Model Analysis Summary

| | Estimate | Standard Error | Confidence Interval |
|---|---|---|---|
| $\alpha$ | −117.025 | 24.737 | {−169.75, −64.2992} |
| $\beta$ | 0.194335 | 0.0149526 | {0.162464, 0.226206} |

| | DF | SS | MS |
|---|---|---|---|
| Model | 2 | $2.13211\times10^7$ | $1.06606\times10^7$ |
| Error | 15 | 439444. | 29296.2 |
| Uncorrected Total | 17 | $2.17606\times10^7$ | |
| Corrected Total | 16 | $1.09158\times10^7$ | |

### Best Fit Function Form

$$D_E(\text{t}) = -117.025\, e^{0.194335\, t} \tag{3}$$

### Model Accuracy

**Determination Coefficient : 0.979806**

### Prediction

**Table 14 . [CAB(EU18 −)]$_T$ versus $D_E$(t) for $T = \{0, 1, 2, \ldots, t\}$**

| Year | t | Observed | Predicted | SE | CILow | CIHigh |
|---|---|---|---|---|---|---|
| 1995 | 0 | 3.73 | -117.025 | 172.94 | -485.637 | 251.588 |
| 1996 | 1 | 10.9607 | -142.127 | 173.429 | -511.782 | 227.528 |
| 1997 | 2 | 19.7353 | -172.614 | 174.021 | -543.531 | 198.304 |
| 1998 | 3 | 4.55701 | -209.64 | 174.722 | -582.051 | 162.772 |
| 1999 | 4 | -66.0264 | -254.608 | 175.53 | -628.742 | 119.526 |
| 2000 | 5 | -184.083 | -309.222 | 176.431 | -685.275 | 66.8309 |
| 2001 | 6 | -287.725 | -375.551 | 177.389 | -753.647 | 2.54479 |
| 2002 | 7 | -390.503 | -456.108 | 178.348 | -836.248 | -75.9685 |
| 2003 | 8 | -500.423 | -553.945 | 179.222 | -935.947 | -171.942 |
| 2004 | 9 | -646.322 | -672.767 | 179.904 | -1056.22 | -289.31 |
| 2005 | 10 | -842.908 | -817.078 | 180.291 | -1201.36 | -432.798 |
| 2006 | 11 | -1115.85 | -992.343 | 180.343 | -1376.74 | -607.951 |
| 2007 | 12 | -1427.83 | -1205.2 | 180.24 | -1589.38 | -821.031 |
| 2008 | 13 | -1747.01 | -1463.72 | 180.675 | -1848.82 | -1078.62 |
| 2009 | 14 | -1924.14 | -1777.7 | 183.421 | -2168.65 | -1386.74 |
| 2010 | 15 | -2152.72 | -2159.02 | 192.15 | -2568.58 | -1749.46 |
| 2011 | 16 | -2331.41 | -2622.13 | 213.092 | -3076.33 | -2167.94 |
| 2012 | 17 | - | -3184.59 | 254.503 | -3727.05 | -2642.13 |
| 2013 | 18 | - | -3867.69 | 324.939 | -4560.28 | -3175.1 |
| 2014 | 19 | - | -4697.32 | 432.474 | -5619.12 | -3775.53 |
| 2015 | 20 | - | -5704.91 | 585.833 | -6953.59 | -4456.24 |

### Graph

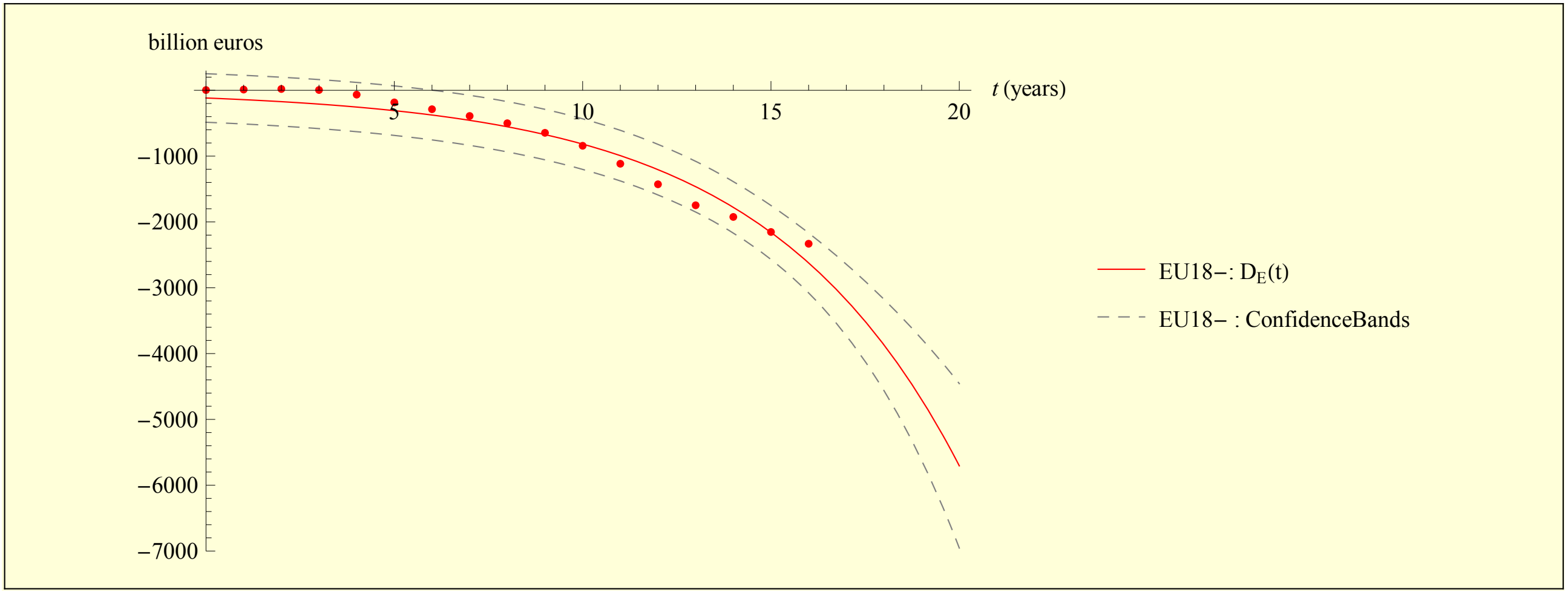

Figure 9: The **[CAB(EU18 −)]$_T$** DataPlot and the **$D_E$(t)** Graph with 95 per cent Confidence Bands

## Graph of the Expected Accumulated Surplus and the Expected Accumulated Deficit Functions for EU

**Figure 10** below, displays jointly the graphs of (**2**) and (**3**) as well as the mirror image of (**2**) with respect to the time axis.

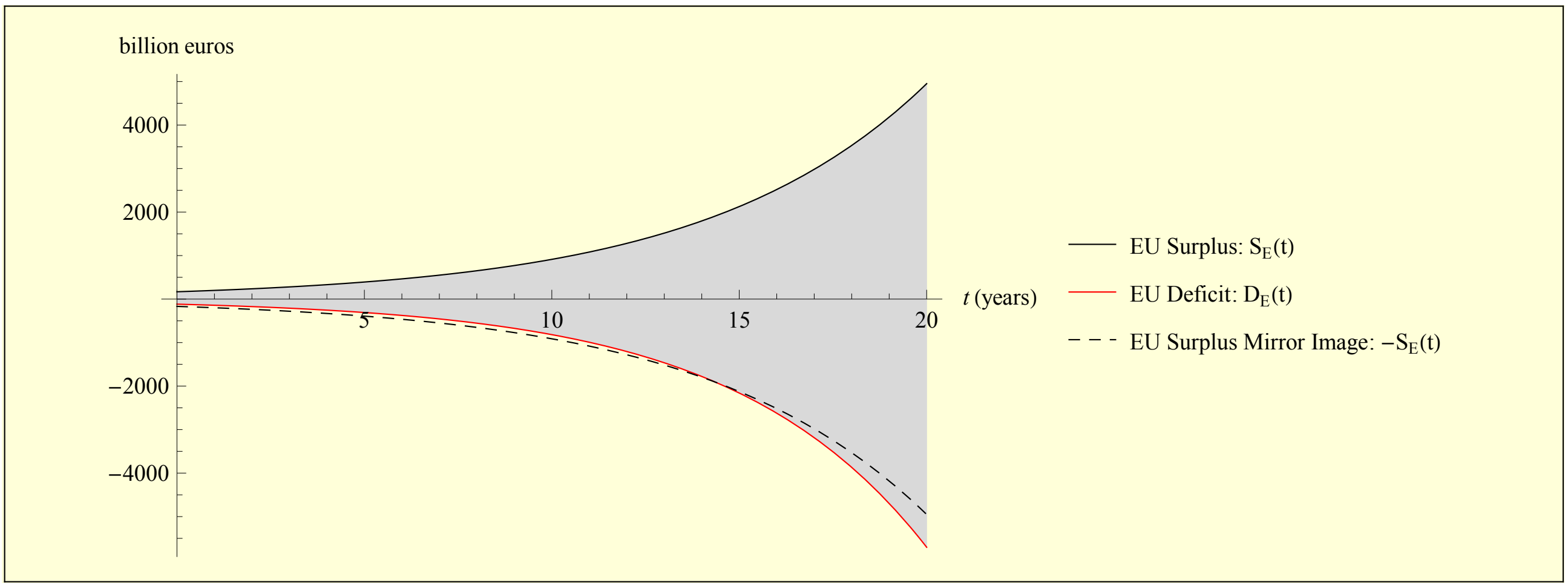

Figure 10: Graphs of the Functions $\pm S_E(t)$ and $D_E(t)$

The asymmetry observed in this figure partly reflects the fact that the deficit of the EU18- countries accumulates at a rate higher than the surplus accumulation rate of the EU9+ countries. Moreover, it must be noticed that the current size of the accumulated deficit is (slightly) larger that the current size of the accumulated surplus and their difference tends to increase with time.

# Curve Fitting the Surplus Accumulation of the Eurozone7+ Countries

### Statistical Model Analysis Summary

| | Estimate | Standard Error | Confidence Interval |
|---|---|---|---|
| $\alpha$ | 136.143 | 19.8392 | {93.8573, 178.43} |
| $\beta$ | 0.171216 | 0.0105005 | {0.148834, 0.193597} |

| | DF | SS | MS |
|---|---|---|---|
| Model | 2 | $1.52695\times10^{7}$ | $7.63473\times10^{6}$ |
| Error | 15 | 191519. | 12767.9 |
| Uncorrected Total | 17 | $1.5461\times10^{7}$ | |
| Corrected Total | 16 | $6.58607\times10^{6}$ | |

### Best Fit Function Form

$$S_\epsilon(t) = 136.143\, e^{0.171216\, t} \tag{4}$$

### Model Accuracy

**Determination Coefficient : 0.987613**

### Prediction

**Table 15 . [CAB(Eurozone7 +)]$_T$ versus $S_\epsilon$(t) for $T$ = {0, 1, 2, ..., $t$}**

| Year | t | Observed | Predicted | SE | CILow | CIHigh |
|---|---|---|---|---|---|---|
| 1995 | 0 | 18.0282 | 136.143 | 114.724 | -108.384 | 380.671 |
| 1996 | 1 | 50.7502 | 161.568 | 115.095 | -83.7511 | 406.887 |
| 1997 | 2 | 112.503 | 191.74 | 115.517 | -54.4796 | 437.959 |
| 1998 | 3 | 163.955 | 227.547 | 115.987 | -19.6728 | 474.766 |
| 1999 | 4 | 205.504 | 270.04 | 116.491 | 21.7455 | 518.335 |
| 2000 | 5 | 218.189 | 320.469 | 117.011 | 71.0657 | 569.873 |
| 2001 | 6 | 276.06 | 380.316 | 117.518 | 129.833 | 630.799 |
| 2002 | 7 | 378.999 | 451.338 | 117.971 | 199.889 | 702.788 |
| 2003 | 8 | 474.677 | 535.624 | 118.323 | 283.424 | 787.825 |
| 2004 | 9 | 650.674 | 635.65 | 118.529 | 383.011 | 888.289 |
| 2005 | 10 | 813.817 | 754.356 | 118.565 | 501.64 | 1007.07 |
| 2006 | 11 | 1023.38 | 895.229 | 118.475 | 642.707 | 1147.75 |
| 2007 | 12 | 1250.03 | 1062.41 | 118.453 | 809.934 | 1314.89 |
| 2008 | 13 | 1412.3 | 1260.81 | 118.998 | 1007.17 | 1514.45 |
| 2009 | 14 | 1566.07 | 1496.26 | 121.157 | 1238.02 | 1754.5 |
| 2010 | 15 | 1752.63 | 1775.69 | 126.823 | 1505.37 | 2046. |
| 2011 | 16 | 1915.49 | 2107.29 | 138.914 | 1811.2 | 2403.38 |
| 2012 | 17 | - | 2500.82 | 161.123 | 2157.39 | 2844.24 |
| 2013 | 18 | - | 2967.84 | 197.281 | 2547.34 | 3388.33 |
| 2014 | 19 | - | 3522.07 | 250.995 | 2987.08 | 4057.05 |
| 2015 | 20 | - | 4179.8 | 325.937 | 3485.08 | 4874.52 |

### Graph

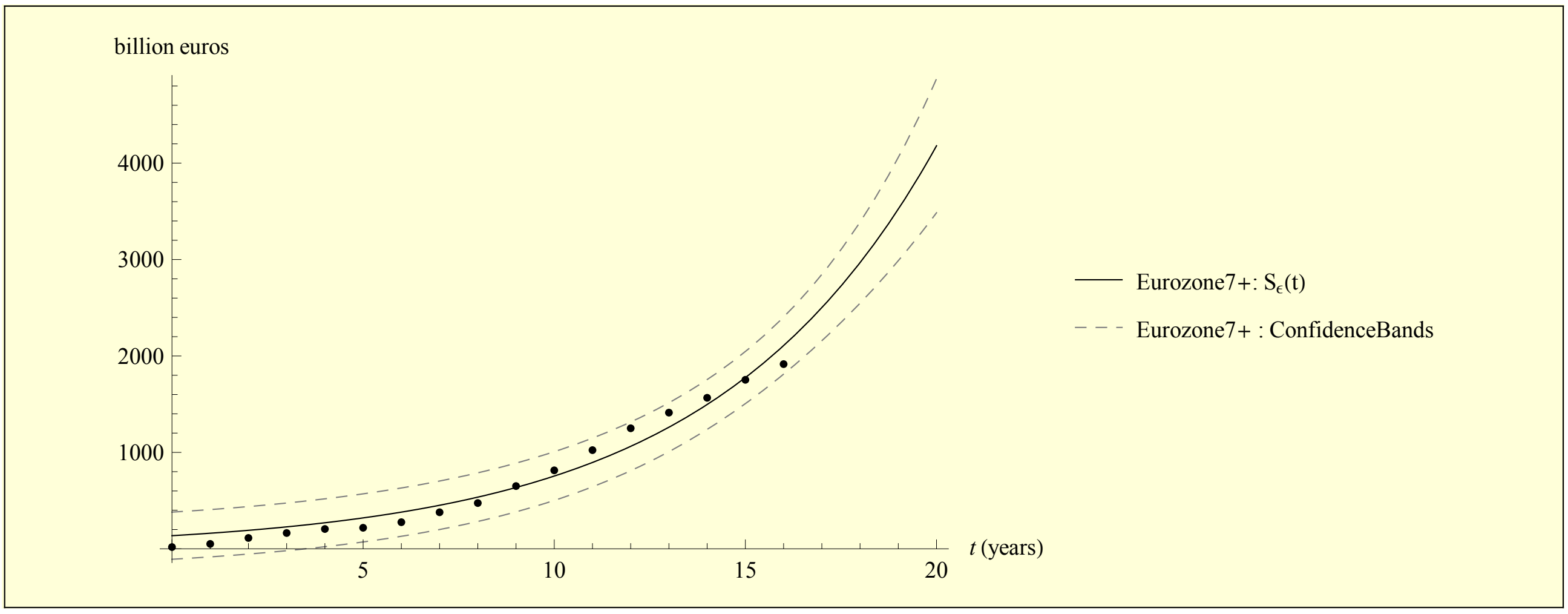

Figure 11: The $[\textbf{CAB}(\textbf{Eurozone7}+)]_T$ DataPlot and the $S_\epsilon(t)$ Graph with 95 per cent Confidence Bands

## Curve Fitting the Deficit Accumulation of the Eurozone10- Countries

### Statistical Model Analysis Summary

| | Estimate | Standard Error | Confidence Interval |
|---|---|---|---|
| $\alpha$ | −36.5652 | 11.9553 | {−62.0474, −11.083} |
| $\beta$ | 0.233702 | 0.0225797 | {0.185574, 0.281829} |

| | DF | SS | MS |
|---|---|---|---|
| Model | 2 | $6.33419\times10^6$ | $3.16709\times10^6$ |
| Error | 15 | 212782. | 14185.4 |
| Uncorrected Total | 17 | $6.54697\times10^6$ | |
| Corrected Total | 16 | $3.97327\times10^6$ | |

### Best Fit Function Form

$$D_\epsilon(t) = -36.5652\, e^{0.233702\, t} \quad (5)$$

### Model Accuracy

**Determination Coefficient : 0.967499**

### Prediction

**Table 16 . $[\text{CAB(Eurozone10} -)]_T$ versus $D_\epsilon$(t) for $T$ = {0, 1, 2, ..., $t$}**

| Year | t | Observed | Predicted | SE | CILow | CIHigh |
|---|---|---|---|---|---|---|
| 1995 | 0 | 18.5691 | -36.5652 | 119.701 | -291.702 | 218.572 |
| 1996 | 1 | 43.5421 | -46.1917 | 119.931 | -301.819 | 209.435 |
| 1997 | 2 | 65.653 | -58.3525 | 120.237 | -314.631 | 197.926 |
| 1998 | 3 | 69.3626 | -73.7148 | 120.636 | -330.845 | 183.415 |
| 1999 | 4 | 51.4743 | -93.1215 | 121.146 | -351.339 | 165.096 |
| 2000 | 5 | -3.17516 | -117.637 | 121.78 | -377.205 | 141.93 |
| 2001 | 6 | -53.8405 | -148.607 | 122.539 | -409.793 | 112.578 |
| 2002 | 7 | -109.068 | -187.731 | 123.407 | -450.766 | 75.304 |
| 2003 | 8 | -170.94 | -237.154 | 124.336 | -502.171 | 27.8624 |
| 2004 | 9 | -248.279 | -299.589 | 125.241 | -566.535 | -32.6442 |
| 2005 | 10 | -370.872 | -378.462 | 125.992 | -647.007 | -109.916 |
| 2006 | 11 | -537.015 | -478.098 | 126.443 | -747.605 | -208.591 |
| 2007 | 12 | -731.758 | -603.966 | 126.531 | -873.66 | -334.272 |
| 2008 | 13 | -959.248 | -762.97 | 126.54 | -1032.68 | -493.257 |
| 2009 | 14 | -1091.57 | -963.836 | 127.737 | -1236.1 | -691.571 |
| 2010 | 15 | -1235.39 | -1217.58 | 133.593 | -1502.33 | -932.835 |
| 2011 | 16 | -1352.05 | -1538.13 | 151.185 | -1860.38 | -1215.89 |
| 2012 | 17 | - | -1943.07 | 190.753 | -2349.65 | -1536.49 |
| 2013 | 18 | - | -2454.62 | 262.878 | -3014.93 | -1894.31 |
| 2014 | 19 | - | -3100.84 | 377.803 | -3906.11 | -2295.57 |
| 2015 | 20 | - | -3917.19 | 547.959 | -5085.14 | -2749.25 |

### Graph

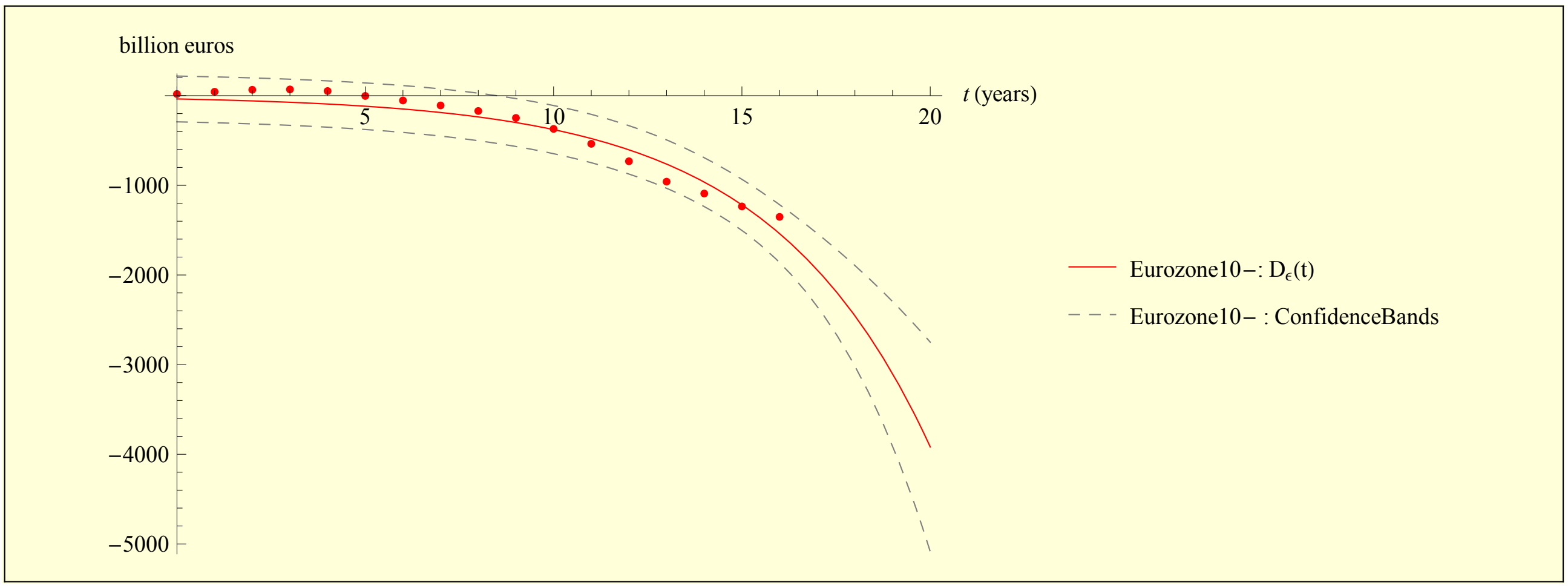

Figure 12:The $[\mathbf{CAB(Eurozone10-)}]_T$ DataPlot and the $D_\epsilon(t)$ Graph with 95 per cent Confidence Bands

## Graph of the Expected Accumulated Surplus and the Expected Accumulated Deficit Functions for the Eurozone

**Figure 13** below, displays jointly the graphs of (**4**) and (**5**) as well as the mirror image of (**4**) with respect to the time axis.

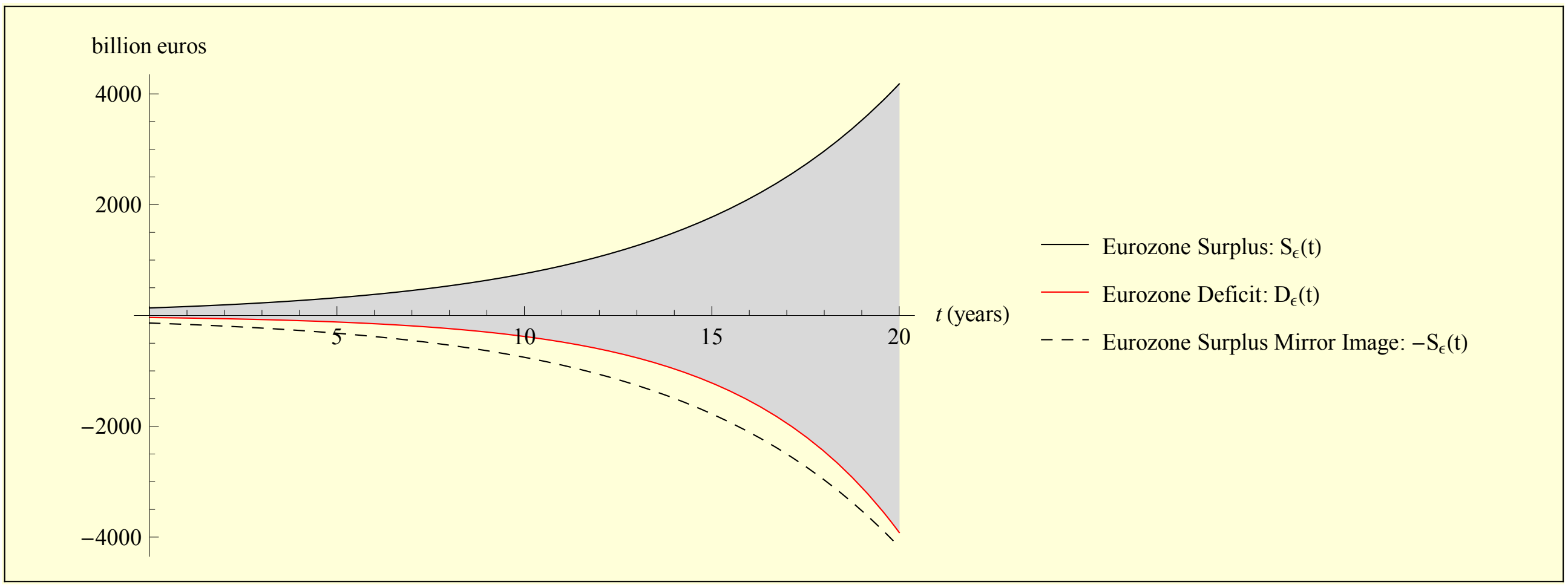

Figure 13: Graphs of the functions $-S_\epsilon(t)$ and $D_\epsilon(t)$

The asymmetry observed in this figure partly reflects the fact that the deficit of the EU10- countries accumulates at a rate higher than the surplus accumulation rate of the EU7+ countries. Moreover, it must be noticed that the current size of the accumulated deficit is smaller than the current size of the accumulated surplus and their difference tends to decrease with time.

# 9. The Surplus-Deficit Gap Functions

## Definitions

The accumulated surplus-accumulated deficit gap for EU, or the *EU surplus-deficit gap* in short, has to be defined as

$$f_E(t) \equiv S_E(t) - |D_E(t)| \qquad (6)$$

Similarly, the accumulated surplus-accumulated deficit gap for the Eurozone, or the *Eurozone surplus-deficit gap* in short, has to be defined as

$$f_\epsilon(t) \equiv S_\epsilon(t) - |D_\epsilon(t)| \qquad (7)$$

## Concativity

Firstly, note that

$$\lim_{t\to-\infty}[f_E(t)] = \lim_{t\to-\infty}[f_\epsilon(t)] = 0 \qquad (8)$$

and that

$$\lim_{t\to+\infty}[f_E(t)] = \lim_{t\to-\infty}[f_\epsilon(t)] = -\infty \qquad (9)$$

Denoting the first and second time-derivatives of the $f$ functions as $f_t$ and $f_{tt}$ respectively, we may summarize the concativity features of the gap functions as follows

| Functions | $signf_{tt}$ | $f_{tt}=0$ | $signf_{tt}$ | $signf_t$ | $f_t=0$ | $signf_t$ | signf | f=0 | signf |
|---|---|---|---|---|---|---|---|---|---|
| Domain | $t<t_2$ | $t=t_2$ | $t>t_2$ | $t<t_1$ | $t_1$ | $t>t_1$ | $t<t_0$ | $t=t_0$ | $t>t_0$ |
| $f_\epsilon(t)$ | + | 11.0803 | - | + | 16.0594 | - | + | 21.0384 | - |
| $f_E(t)$ | + | 3.36206 | - | + | 8.87401 | - | + | 14.386 | - |

Thus, the EU surplus-deficit gap function has reached its maximum value at the year 2003. Since then its value decreases at increasingly faster rates. As for the Eurozone surplus-deficit gap function, its maximum value has been reached at the year 2011 and its value ever since also decreases at increasingly faster rates.

The point at which a surplus-deficit gap function becomes zero, may be called critical or ***instability turning point*** since the gap function beyond this point switches sign and the clear-cut distinction between surpluses and deficits ceases to be valid. If this theoretical scheme turns out to be actually correct, then the instability turning point for EU has already occurred at the year 2009 and EU is currently passing through ***a period of increasing instability***. As for the Eurozone, the instability turning point is yet to occur at the year 2016. Currently, the Eurozone may be said to be passing through ***a period of decreasing stability***.

## Graph of the EU and the Eurozone Surplus-Deficit Gap Functions

**Figure 14** shows the graphs of the two gap functions explored above.

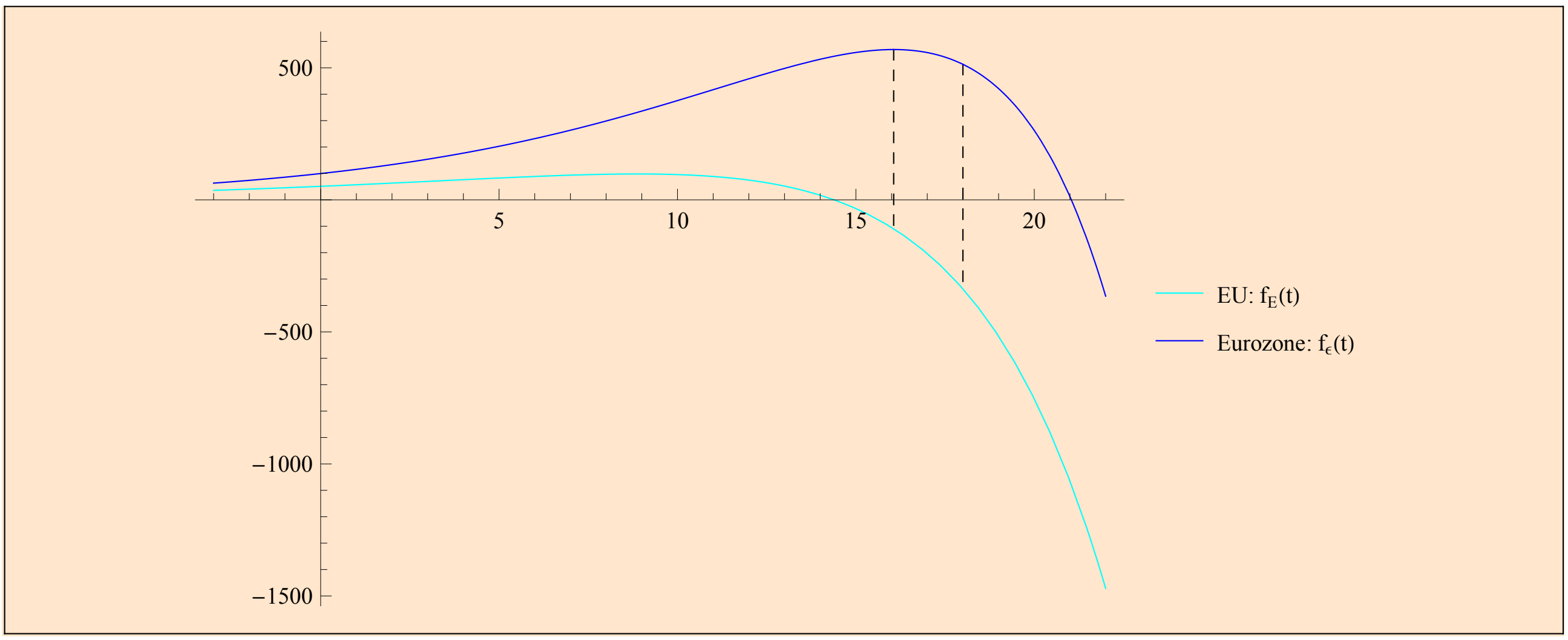

**Figure 14: Graphs of the Surplus-Deficit Gap Functions**

The surplus-deficit gap function characteristics of the Eurozone and of EU for the years 2011 and 2013 (marked by the dashed lines in the figure), are the following:

| Region | $f(t_1[\text{Euro}])$ | $f_t(t_1[\text{Euro}])$ | $f_{tt}(t_1[\text{Euro}])$ | f(18) | $f_t(18)$ | $f_{tt}(18)$ |
|---|---|---|---|---|---|---|
| Eurozone | 569.196 | 0 | -22.7754 | 513.215 | -65.5092 | -47.0613 |
| EU | -109.692 | -85.4918 | -27.4659 | -337.154 | -154.621 | -45.115 |

$t_1$[Euro] refers to to the time when the rate of the Eurozone gap function is zero.

Thus, in dynamical terms, it looks as if both economies are in *free fall* since 2011 under a variable attractive field whose intensity increases rapidly with time. In strictly mathematical terms, it may be said that equation (9) also applies for the rates $f_t$ and $f_{tt}$ of both the Eurozone and the EU economies.

# 10. Uncertainty Time Intervals for the Instability Turning Points

The analytic treatment of the surplus-deficit gap functions that led to the concept of instability turning points deals exclusively with what is expected *on average* for the time changing relation between surpluses and deficits. Consequently, the resulting outlook is deterministic, i.e., too definitive to be actually true.

An alternative, corrective approach is to first supplement the expected surplus and deficit functions with their corresponding standard error estimates, and then set the resulting confidence intervals *in contrast.* This leads almost trivially to uncertainty time intervals enclosing the times when instability turning points are expected to occur. It is self- evident that the two approaches must have identical predictions for the expected occurrence times of the instability turning points.

Let $S(t)$ denote either $S_\epsilon(t)$ or $S_E(t)$ given in (**2**) and (**4**), respectively. Similarly, let $-|D(t)|$ denote either $-|D_\epsilon(t)|$ or $-|D_E(t)|$ found from (**3**) and (**5**), respectively. Also, let $E_S(t)$ and $E_D(t)$ denote the associated standard errors at a level of significance $\alpha$. Then, the $\mathbf{100\,(1-\alpha)\,\%}$ confidence intervals for $S(t)$ and $-|D(t)|$ may be written as

$$S(t) \pm E_S(t) \quad \text{and} \quad -|D(t)| \pm E_D(t)$$

The instability turning point occurs at the time $t_0$ that is found by solving the equation

$$S(t) = -|D(t)|$$

If $t_m$ and $t_M$ are the minimum and maximum solutions among the four distinct solutions of the equations

$$S(t) \pm E_S(t) = S(t_0) \quad \text{and} \quad -|D(t)| \pm E_D(t) = S(t_0), \tag{10}$$

the *uncertainty time interval* for the instability point will then be $t_m \le t \le t_M$, with a *joint* degree of confidence of $\mathbf{100\,(1-\alpha)^2\,\%}$.

# Uncertainty Time Interval for the EU Instability Turning Point

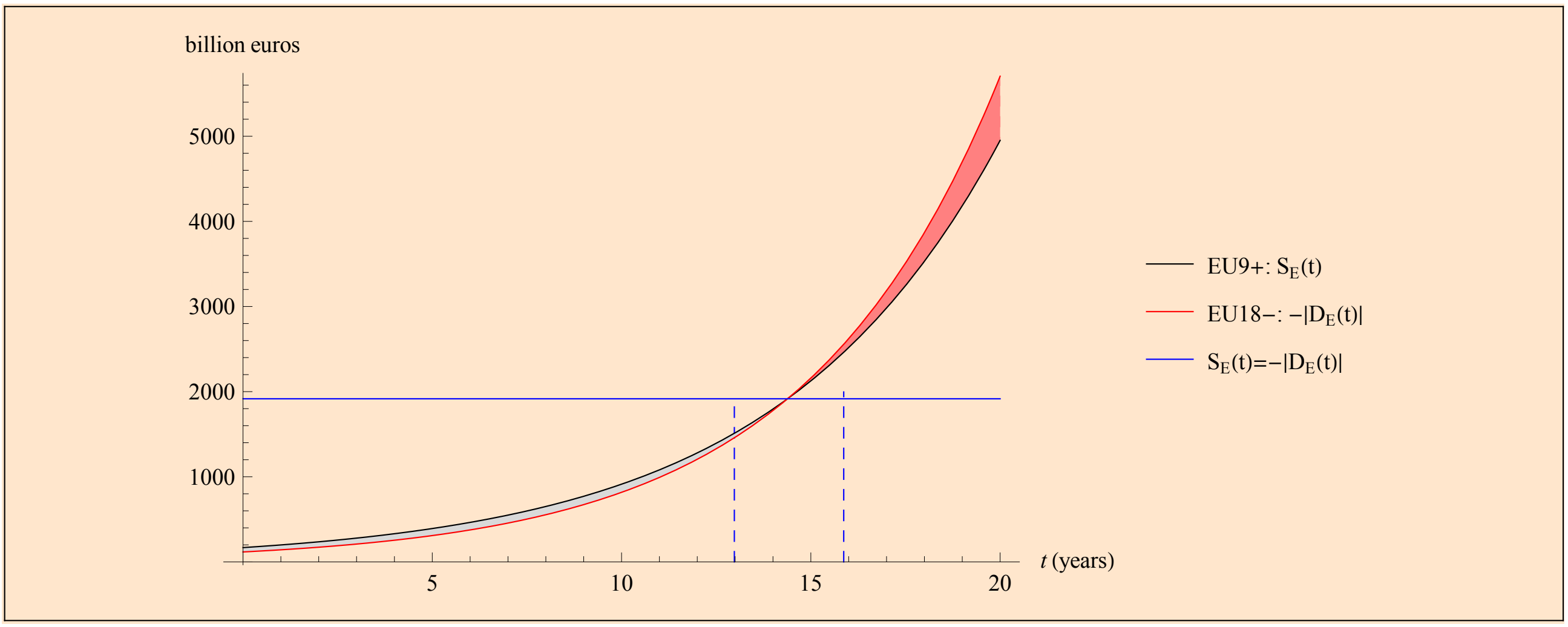

**Figure 15: Graph of the EU Surplus and the Deficit Functions**

**Figure 15** above displays the graphs of the expected EU surplus and deficit functions. The instability point is where the two curves intersect and has the coordinates: **(14.386, 1916.16).** Thus, the instability point was expected to occur at the year 2009. It becomes manifest that the EU surplus curve lies above the EU deficit curve for $t < 14.386$ (the gray shaded region), and below the EU deficit curve for $t > 14.386$ (the pink shaded region). The horizontal blue line shows the height level of the instability point **(1916.16 billion euros).** Finally, the two vertical dashed lines are the graphs of the equations $t = t_m = 12.97947$ and $t = t_M = 15.86883$ marking the endpoints of the uncertainty time interval for the EU instability turning point at a roughly two percent level of significance ($\alpha \simeq 0.02$).

***We may thus conjecture with a ninety eight per cent degree of confidence, that the uncertainty time period for the EU instability turning point to have occurred was any year from 2008 to 2011.***

# Uncertainty Time Interval for the Eurozone Instability Point

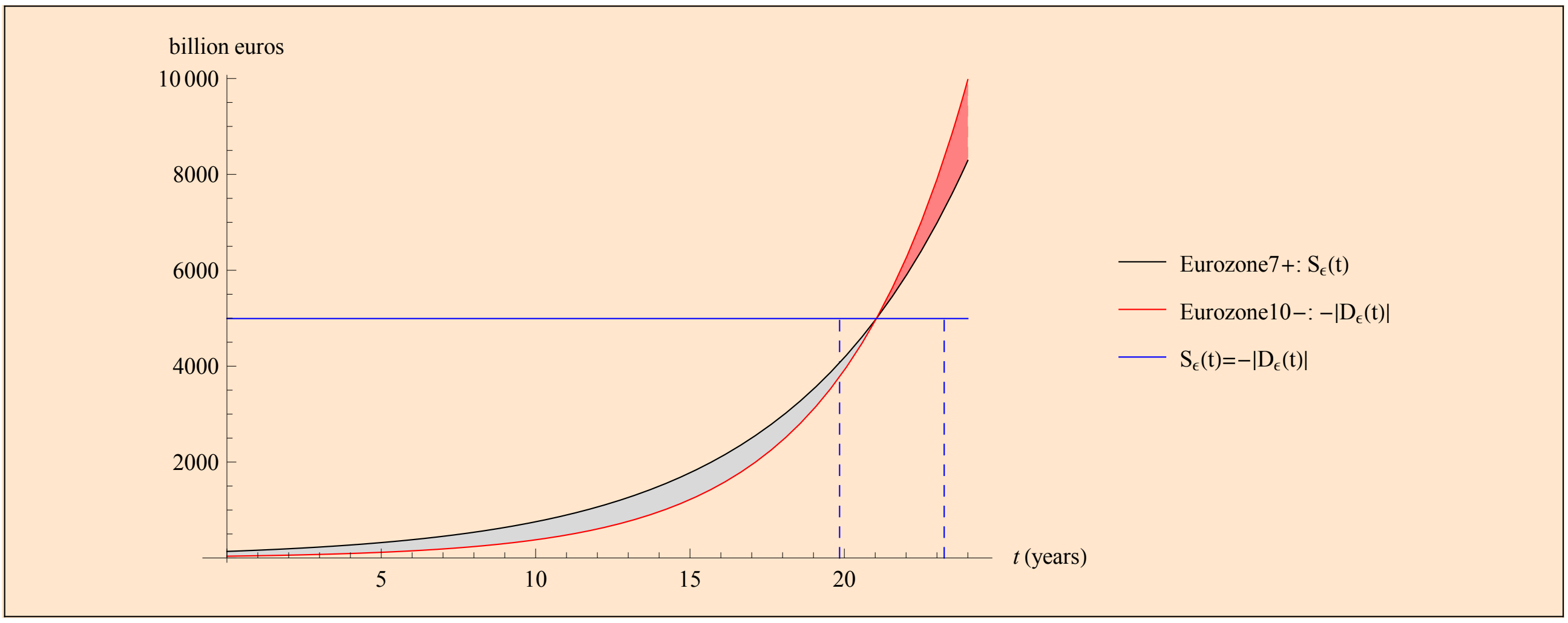

**Figure 16: Graph of the Eurozone Surplus and the Deficit Functions**

**Figure 16** above displays the graphs of the expected Eurozone surplus and deficit functions. The instability turning point is where the two curves intersect and has coordinates: **(21.0384, 4993.13).** Thus, the instability turning point is expected to occur at the year 2016. It becomes manifest that the Eurozone surplus curve lies above the Eurozone deficit curve for $t < 21.0384$ (the gray shaded region), and below the Eurozone deficit curve for $t > 21.0384$ (the pink shaded region). The horizontal blue line shows the height level of the instability point **(4993.13 billion euros).** Finally, the two vertical dashed lines are the graphs of the equations $t = t_m = 19.62491$ and $t = t_M = 23.23408$ marking the endpoints of the uncertainty time interval for the Eurozone instability turning point at a roughly two percent level of significance ($\alpha \simeq 0.02$).

***We may thus conjecture with a ninety eight per cent degree of confidence, that the uncertainty time period for the Eurozone instability turning point to occur is any year from 2015 to 2018.***

# Summary

The first part of this article (Sections 3 to 7) was devoted to an overview of the current account balances and the GDP throughout EU and the Eurozone. By ranking the total current account balances for the time period 1995-2011, a clear-cut distinction was made between surplus and deficit countries and their annual current account balances were then presented and highlighted for various country groupings. Section 8 dealt with the problem of finding functions that best reflect, fit the actual accumulation of surpluses and deficits in EU and in the Eurozone. Asymmetries of a different kind were observed to exist between the surplus and the deficit functions for EU and for the Eurozone. These were explored in Sections 9 and 10 to unveil the concept of the instability turning point of an economy beyond which the accumulation of deficits exceeds the accumulation of surpluses. Finally, at a ninety eight percent degree of confidence, interval estimates were constructed for the occurence time of these instability turning points for the case of EU and for the Eurozone.

# References

1.**Data Source, Eurostat:** http://epp.eurostat.ec.europa.eu/portal/page/portal/government_finance_statistics/data/main_tables.

2.**Editing and Technical Computing Software, *Mathematica*:** http://www.wolfram.com/mathematica/